\documentclass[apj,twocolappendix]{emulateapj}

\usepackage{apjfonts} 
\usepackage{amsmath}
\usepackage{verbatim} 
\usepackage{enumerate}
\usepackage{appendix}
\usepackage{listings}
\usepackage{mathrsfs}

\newcommand{\bjdtdb}{\ensuremath{\rm {BJD_{TDB}}}}

\newcommand{\bjdutc}{\ensuremath{\rm {BJD_{UTC}}}}
\newcommand{\hjdutc}{\ensuremath{\rm {HJD_{UTC}}}}

\newcommand{\tp}{\ensuremath{T_P}}
\newcommand{\tc}{\ensuremath{T_C}}
\newcommand{\ts}{\ensuremath{T_S}}
\newcommand{\ta}{\ensuremath{T_A}}
\newcommand{\td}{\ensuremath{T_D}}
\newcommand{\tla}{\ensuremath{T_{L4}}}
\newcommand{\tlb}{\ensuremath{T_{L5}}}

\newcommand{\fo}{\ensuremath{F_0}}
\newcommand{\dvdt}{\ensuremath{\dot{\gamma}}}

\newcommand{\ms}{\ensuremath{M_*}}
\newcommand{\rs}{\ensuremath{R_*}}
\newcommand{\mpl}{\ensuremath{M_P}}
\newcommand{\rp}{\ensuremath{R_P}}
\newcommand{\fave}{\langle F \rangle}
\newcommand{\fluxcgs}{10$^9$ erg s$^{-1}$ cm$^{-2}$}

\newcommand{\um}{\ensuremath{\mu m}}

\newcommand{\ecosw}{\ensuremath{e\cos{\omega_*}}}
\newcommand{\esinw}{\ensuremath{e\sin{\omega_*}}}
\newcommand{\secosw}{\ensuremath{\sqrt{e}\cos{\omega_*}}}
\newcommand{\sesinw}{\ensuremath{\sqrt{e}\sin{\omega_*}}}
\newcommand{\cosi}{\ensuremath{\cos{i}}}
\newcommand{\sini}{\ensuremath{\sin{i}}}
\newcommand{\cost}{\ensuremath{\cos{\theta_*}}}
\newcommand{\vsini}{\ensuremath{v\sin{i}}}

\newcommand{\logg}{\ensuremath{\log g}}
\newcommand{\logp}{\ensuremath{\log P}}                          
\newcommand{\logk}{\ensuremath{\log K}}

\newcommand{\rhostar}{\ensuremath{\rho_*}}

\newcommand{\feh}{\ensuremath{\left[{\rm Fe}/{\rm H}\right]}}
\newcommand{\teff}{\ensuremath{T_{\rm eff}}}
\newcommand{\rprs}{\ensuremath{R_P/R_*}}

\newcommand{\msun}{\ensuremath{\,{\rm M}_\Sun}}
\newcommand{\rsun}{\ensuremath{\,{\rm R}_\Sun}}
\newcommand{\lsun}{\ensuremath{\,{\rm L}_\Sun}}
\newcommand{\mj}{\ensuremath{\,{\rm M}_{\rm J}}}
\newcommand{\rj}{\ensuremath{\,{\rm R}_{\rm J}}}

\newcommand{\rchisq}{\ensuremath{\chi_\nu^{\,2}}}
\newcommand{\chisq}{\ensuremath{\chi^{\,2}}}
\newcommand{\dchisq}{\ensuremath{\Delta\chi^{\,2}}}
\newcommand{\pchisq}{\ensuremath{P\left(\chisq \right)}}

\newcommand{\kepler}{{\it Kepler}}
\newcommand{\spitzer}{{\it Spitzer}}
\newcommand{\na}{NewA}

\shorttitle{EXOFAST} 
\shortauthors{EASTMAN ET AL\@.}

\begin{document}
\title{EXOFAST: A fast exoplanetary fitting suite in IDL}
\author{Jason Eastman\altaffilmark{1}\altaffilmark{2}\altaffilmark{3}, B. Scott Gaudi\altaffilmark{3}, Eric Agol\altaffilmark{4}}

\altaffiltext{1}
{Las Cumbres Observatory Global Telescope Network,
6740 Cortona Drive,
Suite 102,
Santa Barbara, CA 93117}

\altaffiltext{2}
{Department of Physics Broida Hall,
University of California,
Santa Barbara, CA 93106}

\altaffiltext{3}
{Department of Astronomy, 
The Ohio State University,
140 W.\ 18th Ave., 
Columbus, OH 43210}

\altaffiltext{4}
{Astronomy Department,
Box 351580,
University of Washington,
Seattle, WA 98195;
agol@astro.washington.edu}

\begin{abstract}

We present EXOFAST, a fast, robust suite of routines written in IDL which is designed to fit exoplanetary transits and radial velocity variations simultaneously or separately, and characterize the parameter uncertainties and covariances with a Differential Evolution Markov Chain Monte Carlo method. We describe how our code incorporates both data sets to simultaneously derive stellar parameters along with the transit and RV parameters, resulting in more self-consistent results on an example fit of the discovery data of HAT-P-3b that is well-mixed in under five minutes on a standard desktop computer. We describe in detail how our code works and outline ways in which the code can be extended to include additional effects or generalized for the characterization of other data sets -- including non-planetary data sets. We discuss the pros and cons of several common ways to parameterize eccentricity, highlight a subtle mistake in the implementation of MCMC that could bias the inferred eccentricity of intrinsically circular orbits to significantly non-zero results, discuss a problem with IDL's built-in random number generator in its application to large MCMC fits, and derive a method to analytically fit the linear and quadratic limb darkening coefficients of a planetary transit. Finally, we explain how we achieved improved accuracy and over a factor of 100 improvement in the execution time of the transit model calculation. Our entire source code, along with an easy-to-use online interface for several basic features of our transit and radial velocity fitting, are available online at http://astroutils.astronomy.ohio-state.edu/exofast.

\end{abstract}


\maketitle

\section{Introduction}
In the mere 17 years since the first discovery of a ``Hot Jupiter'' around a main sequence star \citep{mayor95}, the study of exoplanets has exploded into one of the most vibrant and rapidly developing fields in astronomy today. Over 500 exoplanets have been confirmed, and four times that many candidates have been identified \citep{batalha12}. The pace of exoplanet discoveries has consistently increased, with new exoplanet search methods continuously being developed and implemented, and new and surprising classes of planets routinely being uncovered as new regimes of parameter space are explored.  Meanwhile, an enormous amount of effort is being put into developing theories of planet formation and evolution that can encompass the astonishing diversity of planetary systems that has emerged.

The transit method has been at the center of this revolution, not only
because it has expanded the region of parameter space to which we are
sensitive, but more importantly, it can provide a seemingly endless
wealth of information about each planet (see \citealt{winn10a} for a
comprehensive review). For example, precise photometry during the
primary transit can be used to measure the planet radius and orbital
inclination, and when combined with the minimum mass inferred from
radial velocity (RV) studies, yield the true planet mass and average
density, thereby constraining the planet's structure \citep{guillot05,
  sato05, charbonneau06, fortney06}. Photometric observations during
both primary transits and secondary eclipses enable the study of their
atmospheres \citep{charbonneau02, vidal03} and thermal emission
\citep{deming05, charbonneau06, deming06}. Variations in the timing
and shape of the eclipses and transits hint at the existence of other
bodies in the system \citep{miralda02, holman05, agol05, steffen05,
  ford06b, ford07}, constrain orbital evolution due to tides or other
effects \citep{sasselov03, fabrycky08, hellier09}, probe ``weather''
in exoplanet atmospheres \citep{rauscher07}, and provide a probe of
the interior structure and oblateness of exoplanets
\citep{ragozzine09, carter10}, to name a few. In resonant
configurations, even Eris-mass planets may be detectable via such
Transit Timing Variations (TTVs) using current ground-based technology
for favorable systems \citep{carter11}. Further, the projected angle
between the spin axis of the star and the orbit of the planet can be
measured via spectroscopic observations during transit to provide
diagnostic information of the physical processes at work in the
migration of Hot Jupiters \citep{queloz00, winn05, gaudi07,
  triaud10}. Indeed, the combination of radial velocity and transit
data provides the most thorough insights into a planetary system of
any demonstrated planet detection and characterization method to date.

Because of the wealth of information that can be derived from these planets, it is important to carefully consider the best ways to extract such information from the data sets we acquire such that results are limited by the data and can be compared in a manner that is as homogeneous and consistent as possible.

The Markov Chain Monte Carlo method has become a standard tool in exoplanet research \citep[e.g.][]{ford05, gregory05, ford06a, winn10b}, and has recently begun to be replaced with a faster, more elegant flavor: Differential Evolution MCMC, DE-MC \citep[e.g.,][]{braak06,johnson11, carter11b, doyle11}.

Many public codes exist to model transit light curves and/or radial velocities, including TAP \citep{gazak11}, JKTEBOP, \citep{southworth08}, FITSH \citep{pal12}, PHEOBE \citep{prsa05}, VARTOOLS \citep{hartman08}, Nightfall\footnote{http://www.hs.uni-hamburg.de/DE/Ins/Per/Wichmann/Nightfall.html}, PhoS-T \citep{mislis12}, and Systemic \citep{meschiari09}.

The goal of this paper is to present an additional code, EXOFAST, which we believe provides a valuable combination of features not present in any currently-public code: simultaneous and self-consistent radial velocity, transit, and stellar parameter fitting; fast, robust, DE-MC characterization of errors; intuitive outputs, careful attention to realistic priors; non-interactive (easy to pipeline); well documented; and easy to install, use, and customize. Providing a completely general code that can fit any conceivable planetary phenomenon without modification is not practical. Rather than attempt to be comprehensive, our goal was to provide a modular, easily-extensible framework with a relatively straightforward but powerful example implementation for exoplanets that fits a single-planet system which has either or both RV and primary transit data. This framework and example implementation can be adapted to add additional effects as the data are able to constrain them (e.g., TTVs, secondary eclipses), impose different priors, or even analyze completely different problems (e.g., Supernovae, Cepheids).

While IDL is a proprietary language that is generally slower than low-level languages like C and Fortran, we chose to use this language because of the large library of existing code, the ease of development, and the fact that well-written IDL is comparable in speed to higher level languages for most (i.e., non-serial) applications. Of course, an MCMC code is necessarily serial (i.e., one cannot calculate an arbitrary step in the Markov Chain without first calculating the step before it), but the vast majority of the time spent is the model evaluation at each step, which has been carefully vectorized whenever possible. For those unable or unwilling to purchase an IDL license, the GNU Data Language (GDL)\footnote{http://gnudatalanguage.sourceforge.net/} is an open-source compiler that claims full syntax compatibility with code up to IDL version 7.1. Our code does not work out of the box with GDL, but some users have gotten the core features working. Future updates will keep compatibility with GDL in mind.

In \S\ref{sec:overview}, we provide a brief summary of the general problem of fitting data sets, an overview of how MCMC works, and why it is preferred over alternative methods of fitting data and estimating uncertainties. The discussion here and routines cited are completely general to the problem of fitting any model to a data set and properly characterizing the uncertainties -- it is not just applicable to exoplanets or even just astronomy.

Next, we describe our specific procedure to fit RV data (\S\ref{sec:rv}), including a detailed discussion of different ways to parameterize eccentricity (\S\ref{sec:eparam}), and two potentially-common mistakes when using MCMC, both of which can inflate the measured eccentricity of intrinsically circular orbits significantly. We discuss our procedure to fit transit data in \S\ref{sec:transit}, and combined RV and transit data sets simultaneously in \S\ref{sec:rvtran}. Section~\ref{sec:example} walks through an example fit of HAT-P-3b with real, public data to explain how the code works, what it does, and what its outputs are. Our online interfaces to the most useful codes are presented in \S\ref{sec:online}. Along with these online interfaces, all of the source code described here is available online\footnote{http://astroutils.astronomy.ohio-state.edu/exofast/}. Those already familiar with MCMC and the basics of light curve and RV modeling may find it most efficient to begin with the discussion on eccentricity parameterization (\S\ref{sec:eparam}), then skip to \S\ref{sec:rvtran}, where we discuss our unique approach to fitting RV and Transit data simultaneously.

Appendix~\ref{sec:analyticld} demonstrates that the linear and/or quadratic limb darkening coefficients can be fit analytically, which can reduce the dimensionality of a non-linear solver, thereby drastically increasing its speed. Specific improvements to the \citet{mandel02} code to calculate the quadratically limb-darkened flux during transit, including a factor of $\sim$100 improvement in speed that cuts the run time of typical RV and transit fit from an hour to a few minutes, are described in appendix~\ref{sec:occultquad}. Appendix~\ref{sec:random} discusses a problem with IDL's built-in random number generator and provides an alternative at a moderate increase in the overall run time. Appendix~\ref{sec:nege} discusses a way to interpret a negative eccentricity that provides continuous models across the boundary at e=0. Lastly, appendix~\ref{sec:runtime} discusses the execution time and identifies areas for future improvement.
\section{Overview of the Problem}
\label{sec:overview}

\subsection{Finding the best fit}
\label{sec:bestfit}

Given a data set, $D$, with uncertainties, $\sigma$, we would like to generate a model, $M$, from a set of model parameters that describe the data. If we assume the uncertainties are Gaussian, then the probability of the data $D$ given the model $M$, or the likelihood, $\mathscr{L}$, is given by

\begin{equation}
\label{eq:likelihood}
 P(D|M) = (2\pi)^{-n/2}\left[\prod_{i=1}^n (\sigma_i^2 + s^2)^{-1/2}
                        \right] \exp{\left[ -\sum_{i=1}^n \frac{(D_i -
                        M_i)^2}{2(\sigma_i^2 + s^2)}\right]}
\end{equation}

\noindent where the subscript $i$ corresponds to each of the $n$ data points and $s$ is an added scatter term. If we assume $s$ is constant for each model, it can be absorbed by the error in each data point, the likelihood simplifies to

\begin{equation}
  \label{eq:likelihoodsimple}
 \mathscr{L} \propto \exp^{-\chisq/2}.
\end{equation}

\noindent where

\begin{equation}
  \label{eq:chi2}
  \chi^2 = \sum_{i=1}^n\left(\frac{D_i-M_i}{\sigma_i}\right)^2
\end{equation}

Therefore, assuming fixed uncertainties, finding the maximum likelihood is equivalent to finding the model with the lowest \chisq. When the model is linear (i.e., can be written as a simple linear combination of known quantities with unknown coefficients), the \chisq \ can be minimized analytically and exactly to find each of the coefficients (i.e., the parameters of the model) \citep[see][]{gould03}.

However, when the model is non-linear, such as for transits and radial velocities, we must determine the best fit parameters which minimize \chisq \ numerically. Unfortunately, there are no generic algorithms to minimize the \chisq \ for a global parameter space -- often, various tricks are required that are specific to the particular problem at hand. We will discuss the tricks specific to exoplanets in \S\ref{sec:rv} and \S\ref{sec:transit} that we use to restrict the region of parameter space close to the global minimum. Once we identify this region, there are many routines that can robustly find a local minimum. {\tt AMOEBA} is a popular non-linear solver that uses the downhill simplex method to find local minima \citep{nelder65}. Given a starting point and stepping scale (which is approximately the range of parameter space it will consider), {\tt AMOEBA} will crawl through parameter space to find the minimum, using the \chisq \ at each step to determine its next step. This routine is very robust at finding local minima.

IDL comes with its own built-in {\tt AMOEBA} routine, but we discovered a bug that truncates the stepping scale to floating point precision, regardless of the data type initially given. This is detrimental when fitting parameters that require double precision (e.g., Julian Day), since the model will simply oscillate about the minimum and not converge. We provide a debugged version of this code, which now forces all stepping scales to be double precision, as a new code in this suite, {\tt EXOFAST\_AMOEBA}.

Another popular algorithm to find the local \chisq \ minimum is Levenberg-Marquardt (LM) \citep{levenberg44, marquardt63}, which uses numerical derivatives to predict the minimum more precisely after each evaluation, and therefore requires fewer evaluations of the \chisq \ statistic (which is generally completely dominates the computation time). The downside is that, if the \chisq \ surface is not smooth, the numerical derivatives may be a poor predictor of the minimum and the fit will not be as robust.  \citet{markwardt09} published an extremely versatile and widely-used IDL implementation of the LM algorithm, called {\tt MPFIT}. As expected, we found our debugged version of {\tt AMOEBA} to be more robust than {\tt MPFIT}, routinely finding as good and occasionally better values of \chisq, but it also required about 10 times more model evaluations and was therefore about 10 times slower.

Once we have the best fit, we check the quality of the fit by examining the probability that our model has the \chisq \ that it does. If a model is a good description of the data, and the measurement uncertainties are uncorrelated and properly estimated, then the probability of getting the \chisq \ we do, \pchisq, should be 0.5. In the the limit of infinite degrees of freedom, this is equivalent to saying the \chisq \ per degree of freedom, \rchisq \ is unity. Even with as few as 2 degrees of freedom, the difference between $\pchisq=0.5$ and $\rchisq=1$ is only 20\%, which is already better than we expect this method to be. While we do not wish to imply more accuracy than that by being too precise, we scale the errors such that $\pchisq = 0.5$ because it is precisely correct under the most naive assumption that our uncertainties are Gaussian and uncorrelated.

Thus, if $\pchisq$ deviates from 0.5 by an amount that is significantly larger than expected given the number of degrees of freedom, this implies that either the model does not properly describe the data (e.g., there is signal of another planet that has not been modeled), that the uncertainties have not been properly estimated (e.g., because of unrecognized systematics), or that the data are not Gaussian distributed (e.g., there are large, non-Gaussian outliers). Unfortunately, it is often difficult to distinguish between these possibilities a priori.

If one has reason to believe that the model properly describes the data, but nevertheless finds that the $\pchisq$ differs significantly from 0.5, the natural conclusion is that the uncertainties have been misestimated for the bulk of the data (often underestimated) or that there are a few non-Gaussian outliers. Parameters or parameter uncertainties derived from such data are likely to be biased. In this case, there is a strong motivation to attempt to ``correct'' the data such that $\pchisq \sim 0.5$, thereby producing (hopefully) less biased parameters and parameter uncertainties.

A plausible procedure for ``correcting'' the uncertainties is as follows. First, one can search for and identify strong outliers. If, after eliminating these outliers, one still does not find a satisfactory fit, the next step is to modify the uncertainties.  There are generally three ways in which one can modify the uncertainties to $\pchisq=0.5$: scale the uncertainties by a constant multiplicative factor, add a constant term in quadrature to the uncertainties, or both.  The appropriate method to adopt depends in detail on why the uncertainties were misestimated.  However, if this was known, then it is likely that it would have been corrected in the first place.  As a general rule, if there is an unaccounted systematic that is independent of the signal (e.g., stellar jitter), then the additional uncertainty term should be added in quadrature. If there is an error in the normalization or some calibration systematic (e.g., an error in the gain), it is more appropriate to multiply the uncertainties by a constant factor. The practical difference between these two approaches is usually minimal, but adding uncertainties in quadrature will tend to even out all of the uncertainties, whereas multiplying will preserve the relative uncertainties.

In \citet{lee11}, we used our code to fit radial velocity data for a brown dwarf candidate from the MARVELS collaboration.  We found for this candidate that the \rchisq \ for the native data and uncertainties was considerably larger than unity. We then tested many different permutations of eliminating outliers, scaling uncertainties, and adding uncertainty terms in quadrature, to force $\rchisq=1$, and assessed the effect of these different procedures on the resulting best-fit parameters and uncertainties. We found no statistically significant difference between the various methods of altering the uncertainties. Partly motivated by this experience, our default procedure is to scale the uncertainties by a constant factor and we do not include a jitter term, as is common with RV fits. However, we recognize that this result is unlikely to be generic, and note that it is relatively straightforward to modify this procedure in our routines.

Several alternatives to our method of error scaling have been suggested. It has been proposed to allow the uncertainty scaling \citep[e.g.][]{gregory05} or another uncertainty term to add in quadrature \citep[e.g.][]{ford06a} to vary as a free parameter (e.g., the $s$ term in equation~\ref{eq:likelihood}). More recently, \citet{carter09} suggests a wavelet analysis method to fit the correlated noise component more robustly, and states that treating correlated noise as white noise like we do systematically underestimates the errors. Implementing a wavelet analysis is on our long list of eventual improvements, but in practice, the difference is relatively small as long as the red noise is not dominant (see \citet{carter09} for a discussion).

If we consider multiple independent data sets from different sources, they are likely to have different systematics. Therefore, it is a good idea to fit each data set and scale the uncertainties independently, and then ensure that the resulting parameters are consistent with one another before attempting to combine them. If there are large inconsistencies between the data sets, it is indicative of serious problems with either the model (neglected effects) or data (systematic uncertainties) and a simultaneous fit of all data sets should not be trusted. If they are consistent, we can find the best fit to the combined data set using a local minimum solver starting at the best-fit values of one of the independent data sets.

\subsection{Finding the uncertainties: MCMC}
\label{sec:mcmc}

So far, we have discussed the process of evaluating the probability that a given data set $D$ is described by a given model $M$, $P(D|M)$. However, what we are actually interested in is the probability that a given model $M$ is correct, given our data $D$, $P(M|D)$.  This probability depends not only on $P(D|M)$, but also on our prior beliefs about the probability of a given model $M$, $P(M)$, which are related by Bayes' theorem

\begin{equation}
  \label{eq:bayes}
  P(M|D) = \frac{P(D|M)P(M)}{P(D)}.
\end{equation}

\noindent Here, $P(D)$ is the probability of the data, which is given by the integration of $P(D|M)P(M)$ over the parameter space encompassed by the model $M$. Heuristically, Bayes' theorem can be thought of as providing a rigorous way of incorporating new data to revise initial beliefs.

In principle, one is interested in evaluating $P(M|D)$ for two purposes. First, a comparison of $P(M|D)$ between two different models for the same data set can determine which model is more likely to be correct given the data, models, and priors. Second, for a given model, $P(M|D)$ can be used to determine the relative probability of different parameters of the model, also called the posterior probability density. In our case, where we are only considering different parameters of the same model, $P(D)$ is constant, so we need not explicitly calculate it.

Since there is zero phase space precisely at the best fit, it tells us nothing of the uncertainties of the model parameters. However, by evaluating the posterior probability density, $P(M|D)$, we then determine the uncertainties of the model parameters, e.g., by determining the range of a given parameter that encompasses some set fraction of the probability density. Once the priors, $P(M)$, are specified, the task then becomes evaluating $P(M|D)$. The Markov Chain Monte Carlo technique provides an efficient method for doing this that allows for easy determination of median values, uncertainties, and covariances for the fitted model parameters, in addition to any parameters that can be derived from the model parameters. The MCMC technique is also attractive because it is based on the data {\it as given}, as opposed to bootstrap analyses which use simulated data to evaluate the uncertainties.

We adopt the Metropolis-Hastings algorithm to sample $P(M|D)$. We start with a set of trial parameters, and evaluate \chisq \ for this trial set. We then randomly choose a different set of parameters, and calculate \chisq \ for this set of parameters. The ratio of the likelihood, assuming the errors are constant (i.e., equation~\ref{eq:likelihoodsimple}) for the new set of parameters relative to the initial set is given by

\begin{equation}
  \label{eq:likeratio}
  \mathscr{L}_2/\mathscr{L}_1 = \exp^{[\chi^2(M_1) - \chi^2(M_2)]/2}.
\end{equation}

\noindent We then draw a random number uniformly distributed between 0 and 1. If the random number is greater than the likelihood ratio, the model is rejected and we do not step there. Instead, we duplicate a copy of the previous position in the chain as the current step. If the random number is less than the likelihood ratio, we accept the new model. Note, when $\dchisq < 0$ (i.e., the new model is a better fit), $\mathscr{L}_2/\mathscr{L}_1$ is always greater than 1, and so the step will always be accepted.

We repeat this process, stepping to a new region of parameter space until we have a smooth distribution of values for each parameter. The resultant density of steps is proportional to the posterior probability of each parameter, naturally resulting in a robust estimate of the median value and the 68\% confidence interval. Again, we do not consider the absolute normalization, $P(D)$, which one would need to do in order to consider the relative likelihood of models with a different parameterization.

One of the most problematic aspects of using MCMC is determining an appropriate stepping scale and direction. Using the proper length scale is key to speedy convergence. If the scale is too large, very few trial links will be chosen, and many models will be calculated unnecessarily. If the scale is too small, many links will be accepted but the adjacent links in the chain will be highly correlated with each other, and the resultant chain will not be ``well-mixed,'' discussed below. Similarly, if the chains step in a correlated parameter set (a non-orthogonal direction), the chains are very likely to be rejected because the effective step in the orthogonal space will be too large.

The ideal step mimics the posteriori probability distribution precisely \citep{gelman03}. Of course, this is not known a priori; it is exactly what we are trying to calculate. An elegant solution to this problem is the Differential Evolution Markov Chain method of \citet{braak06}, which runs many chains in parallel (equal to twice the number of free parameters), and uses the difference between the parameter values between two random chains to determine the next step. Since the ensemble of chains should be distributed according to the posterior probability, the difference between two random chains gives us the rough scale and direction of the step (i.e., the covariance matrix among all parameters), automatically taking into account the correlations between parameters within each step and dramatically decreasing the number of links required for the chains to be well-mixed. \citet{braak06} also adds a small, uniform deviate to each step to ensure the whole parameter space can be reached -- otherwise, the steps could be cyclic, depending on the starting positions of each chain, leaving islands of unexplored parameter space. However, we found that, because the dynamic range of the ideal step sizes (i.e., the uncertainties) of different parameters can be arbitrarily large depending on the units of the parameters, adding the same uniform deviate to the steps in each parameter is not general. That is, the log of the period has typical errors of order $10^{-6}$, so adding a random deviate of $10^{-4}$ completely dominates its step size, making the step too large and the chain inefficient. However, adding the same random deviate to the systemic velocity, which has typical errors of order $10$ m/s, is completely negligible and does not serve its purpose to adequately mix the steps.

Instead, we estimate the stepping scale by starting at the best-fit values and then varying each parameter individually until the \dchisq \ is one, in our program {\tt EXOFAST\_GETMCMCSCALE}. Then, we add a uniform deviate equal to a small fraction of that step size (we somewhat-arbitrarily picked 1/10). Even with highly-correlated parameters, this algorithm yields an acceptance rate of 17\% for a large number of parameters -- close to the optimal acceptance rate of $\sim20\%$ \citep{gelman03}. When done this way, we can step in all parameters simultaneously without having to monitor the acceptance rates of each parameter individually because their step sizes are self-adjusting.

In order to determine when our MCMC chains have converged, we roughly follow the guidelines set forth by \citet{ford06a}. Each parameter in each of our chains begins at their best-fit value plus 5 times their corresponding step size (approximately their uncertainty) times a Gaussian-distributed random number. We take steps as described above in all parameters simultaneously, until the chains have converged. We consider the chains to be converged when both the number of independent draws, $T_z$ is greater than 1000 and the Gelman-Rubin statistic, $\hat{R}_{\nu}$ is less than 1.01 for all parameters. The independent draws and Gelman-Rubin statistic are calculated in {\tt EXOFAST\_GELMANRUBIN} and defined by \citet{ford06a}. This test must be passed 6 consecutive times -- after passing these tests the first time, we take 1 percent more steps and check again. If it fails, we restart the convergence test. If it passes, we repeat, taking 2, 3, 4, and finally 5 percent more steps. When all tests have been passed consecutively for all parameters, we consider the chains well-mixed and stop. Finally, we find the first point at which all chains have had a \chisq \ below the median \chisq \ and discard everything before that as the ``burn-in'' \citep[e.g.,][]{tegmark04,knutson09}. This eliminates any bias due to the starting conditions.

Due to the limitations of 32-bit operating systems, a single program (e.g., IDL) cannot allocate more than 2 GB of memory ($\sim$260 million double-precision elements). Given the number of parameters, links, chains, and redundant copies of each that must be stored, it is relatively easy to reach this limitation with a moderately-sized chain before it converges. With a 64-bit machine (and IDL), it is possible to increase the maximum length of each chain dramatically, but managing that volume of data is slow. Fortunately, due to the autocorrelations within the chains, we can ``thin'' them with little penalty in the accuracy of the results or the efficiency of convergence, but with a huge benefit to the manageability of the data set. Therefore, we include an option to thin the chain by a factor of $N_{THIN}$, which will only keep every $N_{THIN}$ link in chain as the Markov chain is calculated. EXOFAST will estimate how many steps will be required for convergence after 5\% of the chain has been calculated. If it is not expected to be well-mixed at the conclusion of program, it will output a warning with a recommended thinning factor. Specifying this thinning factor will help ensure that the final chain is converged, but with the unavoidable consequence of increasing the execution time by roughly a factor of $N_{THIN}$.

The resultant parameter distributions (i.e., the histogram of steps for each parameter) are proportional to the posterior probability of each parameter. We quote the median of the distribution as the final value and 34\% confidence interval on either side as the uncertainty. If there are large covariances or non-Gaussian distributions of parameters, the ensemble of median values, while individually most probable, may be a poor fit to the data. This also means that the quoted values for derived parameters are likely to be mathematically inconsistent (e.g., the median values of $e$ and $\omega_*$ do not exactly imply the median value of \ecosw), but they are statistically self-consistent. The true ``best-fit'' set of parameters can be extracted from the program outputs if desired, but we reiterate that the best-fit has zero phase space associated with it, so this is generally not useful.
\subsection{Priors}
\label{sec:priors}

One of the subtleties of a proper MCMC implementation is the correct choice of priors. We implicitly impose a prior that is uniform in each parameter we step, so we must be careful to consider our choice of parameterization carefully such that it matches our a priori theoretical expectation. Alternatively, we can weight the stepping probability by the Jacobian to transform into our desired parameterzation \citep{ford06a}, or correct our posterior distributions afterward by importance sampling, as long as the particular prior chosen does not preclude viable regions of parameter space. Supporters of the MCMC method argue that all methods have some implied bias, and that MCMC is a good way to make that bias explicit. Ideally, our prior would represent the underlying distribution of that parameter given the selection effects inherent to the sample. In practice, this is a very difficult quantity to determine, and often times, the reason we are doing the measurement in the first place is to determine such broad statistics.

Fortunately, when the data are highly constraining, the prior has little influence on the measured value. However, when the data are not highly-constraining, the prior can have a large impact on the measured values.

\subsection{Hybrid fitting: Non-linear and Linear Parameters}
\label{sec:hybrid}

In principle, it is possible to solve exactly for linear parameters during each step of the Markov chain, which will reduce the dimensionality of Markov chain, making convergence much faster. Unfortunately, we have run our program on the same data set, only changing whether or not we fit some subset of parameters linearly or non-linearly, and the uncertainties in the linear parameters were as much as 10 times smaller when fitted linearly at each step. This is not too surprising, since the uncertainties in a Markov chain come from the distribution of parameters at each step. If we always find the very best set of linear parameters given the particular set of non-linear parameters, we should expect the width of that distribution to be narrower than we would find if they were allowed to step randomly at each step, as they do when they are treated as non-linear parameters.

Even if the fitted parameters are not intrinsic to the problem (i.e., their values and uncertainties do not directly affect any physical property we care to measure), it may be that their covariances with parameters we care about cause us to underestimate their uncertainties. To deal with this, we simply fit all parameters non-linearly. It is likely possible to analytically compensate for this decreased scatter, and therefore recover the speed benefit of analytically fitting many parameters \citep[e.g.][Appendix D]{an02}, but we leave this as a potential future improvement.

We also note that this hybrid fitting is similar to detrending data prior to fitting it, as is common in transit fits. In fact, detrending prior to fitting is significantly worse than this because, rather than underestimating the uncertainties and covariances of detrending parameters, it ignores them altogether. This may lead to, among other things, spurious claims of Transit Timing Variations (TTVs), as the most commonly-removed trend (a linear trend with time or airmass) is highly covariant with the transit time.
\section{Radial Velocity}
\label{sec:rv}
\subsection{Radial Velocity Model}

The gravitational interaction between a host star and orbiting planet results in a Doppler shift of the observed spectrum of the star of the form

\begin{equation}
  \label{eq:rv}
  RV(t) = K\left[\cos\left(\theta(t) + \omega_* \right) + \ecosw\right] + \gamma + \dvdt(t-t_0)
\end{equation}

\noindent where $K$ is the radial velocity semi-amplitude, and is equal to

\begin{equation}
  \label{eq:KP}
  K = \left(\frac{2 \pi G}{P(M_{*} + M_{P})^2}\right)^{1/3}\frac{M_P\sin{i}}{\sqrt{1-e^2}}.
\end{equation}

In equations \ref{eq:rv} and \ref{eq:KP}, $\theta(t)$ is the true anomaly as a function of time, $\omega_*$ is the argument of periastron of the star's orbit measured from the ascending node to its periastron\footnote{Throughout this paper, when we reference the argument of periastron, we refer to the argument of periastron of the star's orbit. Since we measure the radial velocity of the star's orbit, this is the often unspoken -- but not completely universal -- standard in the exoplanet literature. The argument of periastron for the planet, $\omega_P$ differs from $\omega_*$ by $\pi$. As this definition can be somewhat counter-intuitive, we keep the subscript ``*'' to make it explicit.}, $e$ is the orbital eccentricity, $\gamma$ is the systemic velocity (or often just an arbitrary instrumental offset), \dvdt \ is a systematic acceleration either due to an additional body in the system with a period much longer than the span of the observations, or systematics in the data, and $t_0$ is an arbitrary zero point for the slope, which we define to be the mean of the of input times.

The true anomaly, which is the angle between periastron and the planet, measured from the barycenter of the system, is

\begin{equation}
  \label{eq:trueanom}
  \theta(t) = 2\arctan{\left[\sqrt{\frac{1+e}{1-e}}\tan{\left(\frac{E(t)}{2}\right)}\right]}
\end{equation}

\noindent where $E(t)$ is the eccentric anomaly, given by Kepler's equation,

\begin{equation}
  \label{eq:eccenanom}
  M(t) = E(t) - e\sin{(E(t))},
\end{equation}

\noindent as a function of the Mean anomaly, $M(t)$. Unfortunately, this is a transcendental equation for which no analytic solution for $E(t)$ exists. It must be solved numerically for a given $M(t)$. We use {\tt EXOFAST\_KEPLEREQ}, a slightly improved version of a code written by Marc Buie and Joern Wilms\footnote{http://astro.uni-tuebingen.de/software/idl/aitlib/astro/keplereq.html}, which uses the method by \citet{mikkola87}, and in the case of high eccentricities, uses a Newton-Raphson method to refine the eccentric anomaly. Our improvement handles diabolical inputs that prevent convergence when angles differ by slightly more or less than $2\pi$. While these cases are relatively rare, this routine is called hundreds of millions of times during the MCMC chain, and the original version almost always failed without our fix.

The mean anomaly simply describes a uniformly flowing time and can be computed from the period of the orbit, $P$, and the time of periastron passage, \tp:

\begin{equation}
  \label{eq:meananom}
  M(t) = \frac{2\pi}{P}(t-\tp).
\end{equation}

In many instances, the reverse calculation is also required. That is, we would like to know the time, $t$, that the planet will be at a given true anomaly. For instance, there are many special times in an orbit that one may be interested in knowing, like the time of periastron, \tp, the time of primary transit center, \tc\footnote{The use of this terminology does not mean that the object transits -- this is simply the predicted transit center if the inclination were favorable. This may more appropriately be called the time of (inferior) conjunction.}, the time of secondary eclipse center, \ts, the time when the star is at its ascending node \ta \ (when the RV is at a maximum), the time when the star is at its descending node, \td \ (when the RV is at a minimum), or the time that the $L_4$ and $L_5$ star-planet Lagrange points pass in front of the center of the star, \tla \ and \tlb, respectively, which may be a favorable time to look for transiting Trojan planets \citep{ford06b}. Each of these times correspond to a particular value of the true anomaly, given here:

\begin{equation}
 \label{eq:specialtrueanoms}
  \begin{split}
   \theta(\tp) =& 0,\\
   \theta(\tc) =& \pi/2-\omega_*,\\
   \theta(\ts) =& 3\pi/2-\omega_*,\\
   \theta(\ta) =& -\omega_*,\\
   \theta(\td) =& \pi-\omega_*,\\
   \theta(\tla) =& 5\pi/6-\omega_*, \\
   \theta(\tlb) =& \pi/6-\omega_*.\\
  \end{split}
\end{equation}

\noindent Fortunately, this is much easier. We simply invert equation~\ref{eq:trueanom},

\begin{equation}
  \label{eq:eccenanom2}
  E(t) =  2\arctan{\left[\sqrt{\frac{1-e}{1+e}}\tan{\left(\frac{\theta(t)}{2}\right)}\right]}, 
\end{equation}

\noindent plug E(t) into equation~\ref{eq:eccenanom}, and solve equation~\ref{eq:meananom} for $t$. Our routine for the reverse correction, {\tt EXOFAST\_GETPHASE} has keywords that can calculate the phases of each of the true anomalies described in equation~\ref{eq:specialtrueanoms} or for an arbitrary true anomaly.

While the $\omega_*$ is completely degenerate for a planet in a circular orbit, when the orbit is fixed to be circular, we follow the convention that $\omega_* = \pi/2$. This has the virtue that the expected time of conjunction (or transit) occurs at ``periastron'', i.e., $\tp = \tc$.

\subsection{RV Parameterization}
\label{sec:rvparam}

The choice of parameterization is extremely important because we implicitly impose priors that are uniform in each of these parameters. If these priors are not physically motivated, they may introduce biases in the values of the parameters we infer from the MCMC\@. Often, these biases can be corrected, but if the particular parameterization a priori excludes certain regions of parameter space, it cannot. Second, the choice of parameterization is important because highly-correlated parameter sets converge very slowly. Fortunately, this latter concern is largely addressed by the slightly more sophisticated, DE-MC algorithm.

The parameterization we favor for radial velocities is $\log P, \log K, \secosw, \sesinw, T_C, \gamma$, and  \dvdt. This particular parameterization differs from that suggested in \citet{ford06a} in a few key ways. The parameterization of $e$ and $\omega_*$ is discussed in detail at the end of this section. We both use $\log P$ and $\log K$, but the notion of imposing strict bounds outside of what we think is possible on each of these parameters to make them ``normalizable'' or ``proper'' priors (i.e., the integral over the allowed states is finite) is unnecessary and somewhat misleading.  On page 62, \citet{gelman03} state that we can obtain a proper result from an improper prior as long as the posterior distribution (i.e., our parameter distribution) is normalizable, but he warns that such ``distributions must be interpreted with care -- one must always check that the posterior distribution has a finite integral and a sensible form.'' Artificially imposing boundaries outside of what can be reasonably expected does not free us from this responsibility. In either case, if our posterior distribution is unexpected, we must investigate why. Further, by not imposing strict bounds, we simplify the code and ensure that we do not exclude solutions that may actually be allowed. In practice, this should make no difference as long as the bounds chosen were sufficiently conservative so as not to bias the result. If the data provide no constraint, we would find that our posterior distribution was unbounded, but this would be obvious because our chains would not converge.

\citet{ford05} suggested parameterizing a time in the orbit with a mean anomaly, $M_0$, at some arbitrary zero-point, $T_0$. However, $M_0$ is trivially related to \tp,

\begin{equation}
 \tp = T_0 - \frac{M_0 P}{2\pi}.
\end{equation}

\noindent So, modulo comparatively minor covariances with $P$, it is just as covariant with $e$ and $\omega_*$ as \tp \ is. Imagine, when $e$ is very nearly zero, the periastron is poorly defined, and may swing wildly from point to point in the orbit at each step in the chain. This means that \tp, and therefore $M_0$, must randomly swing wildly in the same direction or the model would be out of phase with the data and the step would be rejected. This is the essence of why covariant parameterizations are inefficient.

\citet{ford06a} suggested many other alternative parameterizations to aide convergence, including $\omega_* + M_0$ for low-eccentricity orbits, and $\omega_* + \theta_0$ for high eccentricity orbits, where $\theta_0$ is a an arbitrary zero point in the true anomaly. These approximate the angular position of the star relative to the plane of sky at a given reference time, and thus are an attempt to compensate for the poorly-constrained $\omega_*$ (the angle between the angle on the sky and periastron). Looking back at equation~\ref{eq:specialtrueanoms}, however, we see that each of those special times correspond to a fixed angle along the orbit with respect to the plane of the sky with no approximation whatsoever. Indeed, any true anomaly of the form $C - \omega_*$, where $C$ is a constant, is stationary on the sky. Since all such times are similar, we use \tc \ for its practical uses -- \tc \ is the time we want to look for transits (and is the reason we first considered this family of parameterizations).

In addition to the faster convergence time, there are other major advantages to this parameterization.  \tc \ is usually much better constrained (smaller uncertainties) than \tp, we no longer need to tune the parameterization for each system, as \citet{ford06a} recommends, and \tc \ is a parameter of general interest -- either when the planet transits, or in the case of RV planets, when we may wish to search for transits.

One complication of using \tc \ is that it could take on any value that differs by integer multiples of the period and it would have no effect on the derived model. However, the derived uncertainty of \tc, and its covariance with $P$, is strongly dependent on the choice of the zero-point. It is usually best constrained and least covariant closest to the error-weighted mean of the input times, and that is what we use to run the Markov Chain.

An alternative parameterization, which we do not adopt but has its appeal, is $T_{C,1}$ and $T_{C,2}$, the times of the first and last transit in the data set, instead of \tc \ and $P$. Since we almost always know the number of periods, $N_{Orbits}$, in between with no ambiguity, $P$ can be derived trivially at each step ($P=(T_{C,2} -T_{C,1})/N_{Orbits}$), and unlike \tc \ and $P$, $T_{C,1}$ and $T_{C,2}$ are completely uncorrelated.
\subsection{Radial Velocity Best Fit}
\label{sec:rvbestfit}

Finding the global minimum \chisq \ to a radial velocity data set can be extremely complicated because of the large volume of parameter space with widely-separated local minima. Fortunately, this process can be greatly simplified if the orbit is circular, in which case the \ecosw \ term drops out of equation~\ref{eq:rv} and $\theta$ simplifies to the the mean anomaly, $\theta = M = 2\pi (t-T_{P})/P$. Then, the RV equation can be re-written as

\begin{equation}
  \label{eq:rvcir}
  RV(t) = A\cos\left(\frac{2\pi(t-\tp)}{P}\right) + B\sin\left(\frac{2\pi(t-\tp)}{P}\right) + \gamma + \dvdt(t-t_0),
\end{equation}

\noindent where $A$ and $B$ are arbitrary coefficients. In this formulation, $A$, $B$, $\gamma$, and \dvdt \ are all linearly related to the RV, and so for a given $P$, the values of all other parameters that minimize \chisq \ can be found analytically \citep[e.g.][]{gould03}. Thus, there is only one non-linear parameter, $P$, which can be quickly stepped through while the others are solved analytically at each step. From the best-fit values of $A$ and $B$, the parameters of interest can be determined,

\begin{equation}
  \begin{split}
    K     &= \sqrt{A^2 + B^2} \\
    T_{P} &= \frac{P}{2\pi}(\arctan(A,B) + \pi/2). \\
  \end{split}
\end{equation}

\noindent where $\arctan(A,B)$ represents $\arctan(B/A)$\footnote{Mathematically, the arc tangent ranges from $-\pi/2$ to $\pi/2$, but when the sign of the numerator and denominator are known independently, the precise inverse mapping to the full range $-\pi$ to $\pi$ can be determined. This notation is commonly used in computer programming, and is sometimes called {\tt atan2}.}, and $\pi/2$ comes from the definition that $\omega_* = \pi/2$ for circular orbits. This is the basis for the Lomb-Scargle periodogram \citep{lomb76, scargle82}.

The optimal period spacing requires that we sample finely enough such that the difference in phase between the first and last observation changes by $\lesssim 1/2$ radian for each step $dP$. This is because when the phase changes by $\pi$ radians, the RV will have an opposite sign and give a poor fit to the data. This results in the criteria that $dP \le P^2/(4 \pi T)$, where $T$ is the duration of the observations. Typically, a scan through periods in this way will reveal several peaks in likelihood, due to the planetary orbit and aliases thereof. These provide good starting points for more rigorous, fully non-linear local minimization routines which include eccentricity.  \citet{wright09} suggest a parameterization that reduces the number of non-linear parameters of the full, Keplerian orbit from five per planet to three per planet, which would help robustly fit the full Keplerian solution. However, since we are only concerned with single-planet systems here, this method is not required to quickly and robustly find the best fit, and our concern about hybrid fitting in \S\ref{sec:hybrid} trumps the benefit of this parameterizaton during MCMC fits.

Sometimes, because of aliases, multiple periods will provide similarly good fits. In such cases, each period should be investigated individually, as {\tt AMOEBA} is unable to find minima that are widely separated from the initial values. We do not yet employ the more sophisticated technique outlined by \citet{dawson10} to find the best periods among aliases, as this is usually unnecessary. When there is clearly a unique period associated with the best fit, {\tt AMOEBA} will give us a robust result. Given adequate phase coverage over one or more complete orbits, this method can robustly fit most single-planet systems without any special effort. Those with very high eccentricities or poor phase coverage can also often be fit, but sometimes need slight adjustments to the default period range or number of minima to explore. For example, fitting the data from \citet{winn09b} for HD 80606b ($e\sim0.93$) required searching the 100 likeliest peaks in the Lomb-Scargle Periodogram, whereas the default is 5.

\subsection{Eccentricity}
\label{sec:eparam}

Different groups have chosen to parameterize the eccentricity, $e$, and argument of periastron, $\omega_*$, in several different ways. This is done ostensibly for three different reasons: to apply the appropriate prior, make the chains converge faster, and ease and simplicity in programming. It is unclear what a good, physically-motivated prior for the eccentricity of a planet is, particularly given the complications of tidal circularization. Still, it is commonly assumed that a uniform prior in $e$ is best, and we do the same.

Therefore, we explored the advantages of the most common ways to parameterize the eccentricity, specifically, \ecosw \ and \esinw; $e$ and $\omega_*$ where $0 \leq e < 1$ and $-\pi < \omega_* < \pi$, and \secosw \ and \sesinw. We found subtle differences in the different parameterizations, which are worth further exploration.

In Appendix~\ref{sec:nege}, we discuss a potentially-useful new parameterization, $e$ and $\omega_*$ where $-1 \leq e < 1$ and $\omega_*$ is unbounded. We explain why the models are continuous across the $e=0$ boundary, but ultimately, we found it inferior to our preferred parameterization and do not endorse it for our particular application. 
\subsubsection{Eccentricity Prior}
\label{sec:eprior}

To ensure our prior distribution is what we expect (i.e., uniform), we computed the prior distributions of $e$ for each of the above parameterizations by running a standard Markov Chain, but setting the \chisq \ to 1 as long as the chain made an allowed step. If the chain makes a disallowed step, we set the \chisq \ to infinity. Since this results in a uniform likelihood surface in the allowed region, the posterior distribution is equal to the prior distribution.

\begin{figure}[h]
  \begin{center}
    \includegraphics[width=3.25in]{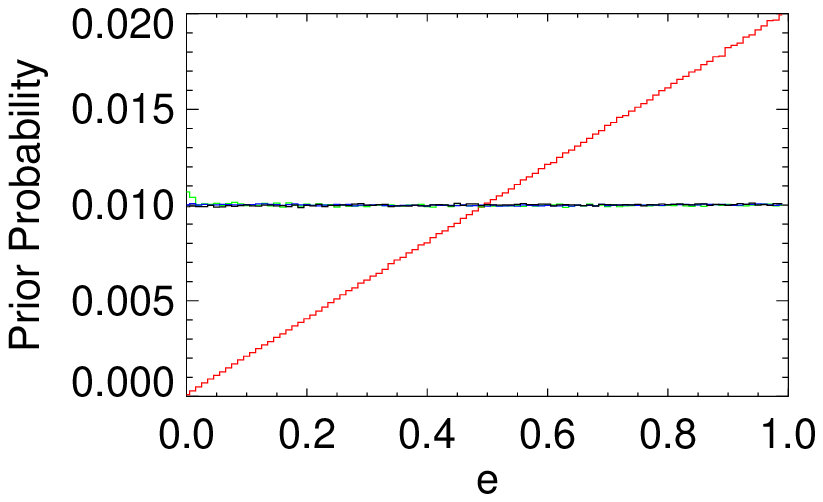} 
    \caption{The prior distributions of $e$ while stepping in \secosw \ and \sesinw \ (black), \ecosw \ and \esinw \ without correcting for the linear prior (red), with correcting for the linear prior (green), and stepping in $e$ and $\omega_*$ directly (blue). We note that the \secosw, \sesinw and $e, \omega_*$ parameterzations correctly reproduce the uniform prior, but the \ecosw, \esinw without correcting for the prior shows a clear linear trend and is obviously wrong. Even correcting for the prior, there is a slight overcorrection at $e=0$.}
    \label{fig:eprior}
  \end{center}
\end{figure}

When parameterizing in \ecosw \ and \esinw, we solve for $e$ and $\omega_*$ at each step, and set the \chisq \ to infinity if $e \geq 1$. As noted by \citet{ford06a}, and shown in figure~\ref{fig:eprior} in red, we clearly see a linear prior in $e$. Ford said that we must correct for the linear prior in $e$ during an MCMC fit by weighting the stepping probability by the Jacobian of the transformation between the parameters in which we step and the parameters we desire to be uniform. In the case of stepping in \ecosw \ and \esinw, the Jacobian to transform to $e$ and $\omega_*$ is $e$, so we must weight the stepping probability by $e_{i-1}/e_{i}$, where the subscript $i$ denotes the current link in the chain. This will preferentially reject steps to higher eccentricity and nearly recovers the uniform prior in $e$, as shown in green in Figure~\ref{fig:eprior}. However, due to the singularity at $e=0$, there is a very slight overcorrection at $e=0$.

Another popular parameterization \citep[e.g.,][]{anderson11} is \secosw \ and \sesinw. Since the Jacobian to $e$ and $\omega_*$ is a constant 1/2, it can be ignored when computing the stepping probability, and it naturally recovers a uniform prior in $e$ and $\omega_*$, as shown in black in Figure \ref{fig:eprior}. This works by essentially taking smaller steps in eccentricity near zero which exactly compensate for the smaller area at $e=0$ in \ecosw, \esinw \ space.

Stepping directly in $e$ and $\omega_*$ obviously results in a uniform prior (shown in blue), where we set the \chisq \ to infinity if our step took us to any of $e < 0$, $e \geq 1$, $\omega_* < -\pi$, or $\omega_* > \pi$.
\subsubsection{Convergence Time}
\label{sec:nsteps}

The primary motivation often cited for using a different parameterization of $e$ is that the convergence time is reduced. To test this, we simulated 100 data sets similar to \citet{ford06a}. In particular, we chose $\gamma = 500$ m/s, $P = 3.223$ days, $\omega_* = 53^\circ$, and $K=50$ m/s, (\dvdt \ was fixed at zero) with 80 evenly spaced data points over a span of 30 periods and Gaussian random uncertainty with a 1-sigma width of $\sqrt{5}$ m/s. The major differences from \citet{ford06a} are that our times were evenly distributed rather than distributed according to the observing times of the California and Carnegie planet search program, and we input Hot-Jupiter-like parameters for those which were not explicitly stated. Neither of these should have an appreciable impact on our comparison.

We fit these 100 simulated data sets as described in \S\ref{sec:rvbestfit} with each of the three eccentricity parameterizations, repeating the procedure for each of the 15 eccentricities listed in Table~\ref{tab:nsteps} (for a total of 4500 fits). The uncertainty in the eccentricity, as found by our Markov Chain, was approximately 0.007 in each case, so our steps in eccentricity roughly correspond to 0 to 10 sigma significant eccentricity. For a more direct comparison with \citet{ford06a}, we also include eccentricities of 0.01, 0.1, 0.5 and 0.8. Each set of 100 simulated RV curves had the same Gaussian ``random'' noise to make the comparison more robust.

We then recorded the number of steps each fit took until the chain was well-mixed, according to the criteria outlined in \S\ref{sec:mcmc}. Table~\ref{tab:nsteps} shows the log of the median number of total accepted steps in all chains until convergence for each of the four parameterizations of eccentricity, as a function of eccentricity, in addition to the best value from all parameterizations proposed by \citet{ford06a}, where applicable. Since the execution time is proportional to the number of steps in the chain (plus small overheads), this is a convenient, computer-independent way of determining how efficient the Markov Chain is. For reference, for the $e=0$ case parameterized as \secosw, \sesinw, the fit took about 15 seconds on a standard desktop computer purchased in 2009.

For moderate eccentricities, \ecosw, \esinw \ was slightly faster than \secosw, \sesinw. Both were nearly twice as fast as $e$, $\omega_*$ at low eccentricities. Even compared to \citet{ford06a}, which uses a significantly more complicated (system-dependent) parameterization, our lowest eccentricity case was about four times faster, and our highest eccentricity case was about 50\% faster. However, at moderate eccentricities, \citet{ford06a} was 2 - 4 times faster. Given the consistency of the number of steps we took as a function of eccentricity and the relatively larger variability of Ford's, some of both our observed benefit and deficiency may be due to random variability of Ford's chains. To give the reader an idea in the variability of our chains, our best chains (as opposed to the median values quoted in the table) at $e=0.1$ and $e=0.5$ took log 4.31 and 4.35 steps, respectively, while our worst chains for $e=0.01$ and $e=0.8$ took log 4.89 and 4.82 steps, respectively.

\begin{deluxetable}{c|cccc}
\tablecolumns{6}
\tablecaption{The log of the number of steps in all chains before convergence for the 3 different eccentricity parameterizations discussed in \S\ref{sec:ebias}, for many values of eccentricity, along with the best values from \citet{ford06a} where applicable. The \ecosw parameterization has been corrected for the linear prior.}
\tablehead{
  \colhead{$e$}          &
  \colhead{\ecosw} &
  \colhead{\secosw}       &
  \colhead{$e,\omega_*$}       &
  \colhead{\citet{ford06a}}
}
\startdata
0.000 & 4.62 & 4.62 & 4.77 &  --  \\
0.007 & 4.58 & 4.62 & 4.76 &  --  \\
0.010 & 4.57 & 4.62 & 4.76 & 5.2  \\
0.014 & 4.57 & 4.64 & 4.76 &  --  \\
0.021 & 4.53 & 4.62 & 4.76 &  --  \\
0.028 & 4.53 & 4.57 & 4.62 &  --  \\
0.035 & 4.53 & 4.57 & 4.57 &  --  \\
0.042 & 4.53 & 4.54 & 4.57 &  --  \\
0.049 & 4.53 & 4.53 & 4.57 &  --  \\
0.056 & 4.53 & 4.53 & 4.57 &  --  \\
0.063 & 4.53 & 4.57 & 4.53 &  --  \\
0.070 & 4.53 & 4.53 & 4.57 &  --  \\
0.100 & 4.53 & 4.53 & 4.53 & 3.9  \\
0.500 & 4.53 & 4.53 & 4.53 & 4.2  \\
0.800 & 4.53 & 4.53 & 4.53 & 4.7
\enddata
\label{tab:nsteps}
\end{deluxetable}

Despite requiring more chains (twice the number of fitted parameters, or 12 in this case), a significant fraction of the advantage we observe relative to \citet{ford06a} (who used 10 chains) is due to the DE-MC algorithm. However, using a standard implementation of the Markov Chain, we were still faster than Ford in the low eccentricity cases due to our use of \tc \ rather than the various combinations of \tp, $M_0$, and $\omega_*$ (see \S\ref{sec:rvparam}).

The final consideration when picking a parameterization are the practical advantages of implementing each. We did not implement the system-dependent parameterizations suggested by \citet{ford06a}. From a practical sense, this is clearly the most difficult, though it may be worth it in the moderate-eccentricity cases, given the results in Table~\ref{tab:nsteps}. If desired, this can be done with relatively minor changes to the code.

Stepping directly in $e$ and $\omega_*$ is intuitive, but dealing with a periodic angular parameter introduces a number of special cases. In particular, the periodic boundary of $\omega_*$ can confuse the AMOEBA algorithm and it may not find the best-fit value. During the MCMC fit, we must be careful to take the modulo whenever it crosses a periodic boundary (i.e., $\pm \pi$) -- otherwise, it would be free to jump between equivalent, widely-separated minima. However, the DE-MC algorithm would fail if the preferred value were near the boundary so that it could draw steps from both sides to determine the step size. Further, we may get unlucky and find that our probability distribution function lies on a boundary or, when it is poorly constrained, it is possible for a significant amount of power to span the entire range of $2\pi$. In either of these cases, the median value, which is required to calculate the convergence criteria and is often used as the final value, would be heavily biased. In order to account for this, we must first center the distribution about the mode, such that the values are within the range mode $\pm \pi$ before we calculate the median. Additionally, our scheme of finding the appropriate step size would fail if the angle is so poorly-constrained that no value of $\omega_*$ produced a $\dchisq=1$.

Stepping in \ecosw \ and \esinw \ eliminates this complicating angular value during the Markov chain, but introduces an arguably more complex requirement to deal with the Jacobian. The \chisq \ routine is required to return a determinant (even if it is 1 for no transformation) to make the priors more obvious to the end-user. After this, the MCMC routine handles the determinant weighting transparently in order to transform to the desired prior. So with our code, using a Jacobian is trivial to deal with. However, as seen in Figure~\ref{fig:eprior}, this does not completely correct the prior.

Stepping in \secosw \ and \sesinw \ eliminates the Jacobian and frees us from most of the burden of dealing with angular parameters, and is therefore practically the simplest. Additionally, it is comparable or faster than other parameterizations, and recovers a precisely uniform prior in $e$. For these reasons, we use it in EXOFAST\@.
\subsection{Eccentricity Bias}
\label{sec:ebias}

It has long been understood that there is a bias against low eccentricities in binary systems. Such a bias is extremely important to understand, as many of the observed systems are expected to be tidally circularized. If they are not, it has profound implications for our understanding of tides (and the tidal Q factor), the existence of additional bodies in the system which may be perturbing their orbits, and the formation and evolution of planets as a whole.

\subsubsection{Lucy-Sweeney bias}
\label{sec:lucysweeney}

The bias against $e=0$ that most people are aware of was first quantified by \citet{lucy71} in the case of binary stars, and is due to the fact that there is zero phase space at exactly $e=0$, and therefore observational uncertainties will produce a best fit that is biased toward a positive value, even for intrinsically circular orbits. They say that, in order to measure a non-zero eccentricity with a 95\% confidence, one must measure a result of $e>2.45\sigma_e$, rather than the naively-expected $e > 2\sigma_e$, where $\sigma_e$ is the standard deviation of the eccentricities.

\begin{figure}[h]
  \begin{center}
    \includegraphics[width=3.25in]{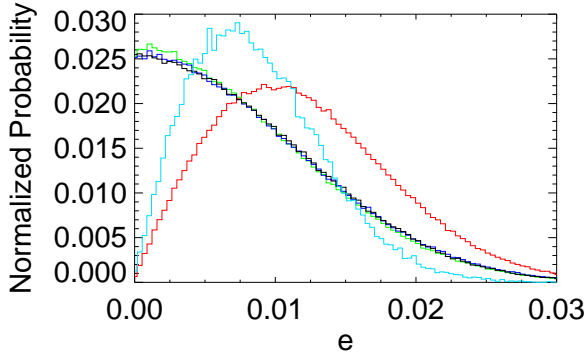} 
    \caption{The combined eccentricity PDFs for all 100 intrinsically circular orbits with the different parameterizations (legend same as figure~\ref{fig:eprior}), plus the distribution of best-fit values from 100,000 amoeba fits (cyan), demonstrating the problem with bootstrap analyses of bounded parameters, the linear prior, and the very slight problem with the corrected \ecosw, \esinw \ paramterization.}
    \label{fig:edist}
  \end{center}
\end{figure}

We simulated 100,000 data sets of intrinsically circular orbits with different noise like those described in \S\ref{sec:nsteps}, and found the best-fit eccentricity using AMOEBA for each. The resultant histogram of best-fit eccentricities is shown in figure~\ref{fig:edist} in cyan. This is the eccentricity distribution one would measure from a bootstrap analysis, similar to that described in \citet{laughlin05}, and clearly shows this deficit at $e=0$.

For comparison, we also plot the combined PDFs of the 100 trials of MCMC fits described in \S\ref{sec:nsteps} for each of the parameterizations of eccentricity. We clearly see the problem of using a linear prior (red). The other distributions look similar, but a close inspection shows the \ecosw, \esinw \ distribution is slightly biased high at $e=0$ because of the slight bias in the prior distribution. Fortunately, the difference between the parameterizations is negligible relative to the width of the Gaussian, so it is of little practical importance.

Now a huge advantage of MCMC becomes apparent: instead of looking at simulated permutations of the data and finding the best fit (which has zero likelihood because it is infinitesimally small) like a bootstrap or prayer bead analysis does, MCMC considers the data as is. It is clear that the PDFs from the MCMC do not suffer from this bias to nearly the same extent, but there is still the matter of how to summarize such a non-Gaussian distribution. Obviously, the standard method of quoting a median value and a 68\% confidence interval provides a misleading, marginally-significant, non-zero eccentricity -- even the absolute value of a Gaussian peaked at zero has a median value of 0.67 sigma. Instead, we could fit a 3-parameter Gaussian to the PDF (normalization, zero point, and width) and use the zero point as the likely value and the width as its uncertainty. In principle, we can actually infer a negative eccentricity, dramatically reducing the Lucy-Sweeney bias. Alternatively, we can quote an upper limit which does not assume the PDF is Gaussian, but does not give us a likely value or uncertainty. In the end, however, there is no substitute for a visual inspection of the PDF\@. By default, EXOFAST simply quotes the median values and 68\% confidence interval as with all other parameters. Therefore, it is up to the user to inspect the PDF and assess the significance of the output eccentricity. Recently, \citet{lucy12} introduced a new Bayesian method to evaluate the robustness of a measured eccentricity, which may help.

\begin{figure}
  \begin{center}
    \includegraphics[width=3.25in]{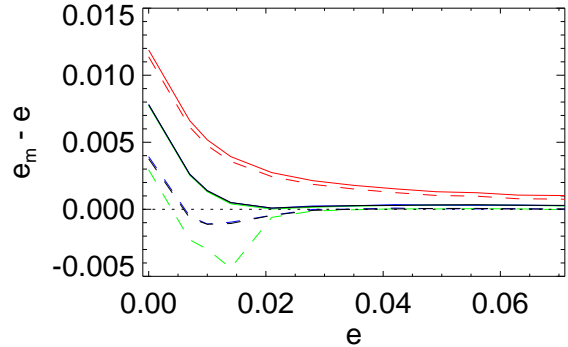} 
    \caption{The difference between the measured and intrinsic eccentricity as a function of the intrinsic eccentricity, using the same parameterizations as before (see Figure~\ref{fig:eprior} for legend). The solid lines are the typical median value, while the dashed lines are the zero point of a 3-parameter gaussian fit to the PDF\@. We clearly see the effect of the Lucy-Sweeney bias at low eccentricities, and the additional bias due to the linear prior for all eccentricities. The values plotted are the average of the median/fitted values for all 100 fits of simulated data. The statistical uncertainties for each of the 100 fits was roughly 0.007 at each point. The small, systematic offset between the measured and intrinsic eccentricities at high eccentricites for all methods is within the uncertainties given we only ran 100 trials.}
    \label{fig:ebias}
  \end{center}
\end{figure}

To see how this bias evolves with the significance of eccentricity, we calculated both the median and fitted values of the eccentricity of each of the 100 fits and averaged them together in order to reduce the statistical uncertainty and flesh out the bias.  This mean of medians is plotted as a function of intrinsic eccentricity in figure~\ref{fig:ebias} in solid lines, and the fitted Gaussian zero-points are shown as dashed lines for each of the the parameterizations described above. A non-biased result would fall on the dotted line. As expected from the Lucy-Sweeney bias, we over-estimate the eccentricity of orbits with intrinsically small eccentricities. The uncertainty in each value is roughly 0.007.

The maximum measured eccentricity deviation out the 100 trials was slightly more than $3\sigma$ -- slightly more than we would expect by chance. In the real world with real data, systematic uncertainties, unmodeled effects, or outliers will all tend to make the eccentricity appear even larger than this data set simulated with white noise.

\citet{laughlin05} pointed out this bias in the case of planetary systems, though they incorrectly called it a Lutz-Kelker bias \citep{lutz73}. The Lutz-Kelker bias is actually a volume effect described in the case of trigonometric parallaxes. While parallax is a positive-definite parameter which suffers from a Lucy-Sweeney bias too, the Lutz-Kelker bias exists at all values of parallaxes. It states that, due to observational uncertainties and the fact that the number density of stars is larger for those with a smaller parallax, more stars tend to have a true parallax that is smaller than what is measured.
\subsubsection{Linear Prior}

As shown in \S\ref{sec:eprior}, if we step in \ecosw \ and \esinw, we introduce a linear prior. This prior is clearly not supported by the observations of short-period planets to date, so failing to correct for it will therefore lead to a significant (up to $\sim5\sigma$) bias in the inferred eccentricity, as shown in Figure~\ref{fig:ebias} in red. Since there is very little difference in the convergence time, many may find it easier to step in \secosw \ and \sesinw, as we do.
\subsubsection{Metropolis-Hastings algorithm}
Another bias comes from an incorrect, but potentially-common, implementation of the Metropolis-Hastings algorithm. As we discussed in \S\ref{sec:mcmc}, when we reject a step, we must make a copy of the previous step in its place. Since the acceptance rate is ideally around 20\%, we will end up with each step copied, on average, five times. This algorithm is somewhat unintuitive: we might think we would end up with huge spikes in the parameter distributions where the chain got stuck (fortunately, not true), or that making 5 copies of each step is wasteful (and it is; see the discussion on thinning in \S\ref{sec:mcmc}). Instead, we might think that we should not copy the previous step. The effect of making this mistake is minimal for unbounded parameters, which makes it difficult to identify with typical sanity checks. However, this misleading intuition can introduce a significant bias that guides fits away from any hard boundary, such as $e=0$. Unfortunately, an obsolete version of an MCMC code which was distributed with the IDL astronomy library made this exact mistake\footnote{The current version, distributed with the most recent IDL Astronomy library, corrects this error, but the old version is still readily available on the web, six years after it was fixed}. Those who use this code, model their own codes from it, or make the same intuitive mistake will all suffer from this bias.

A more subtle way to make essentially the same error is in the boundary handling. The proper meaning of a boundary (e.g., $e = 0$) is that the model's likelihood is zero (or the \chisq \ is infinite) beyond it. However, we cannot a priori restrict the model to step only in bounds. While it may seem more efficient not to allow the Markov chain to step out of bounds, we must allow the Markov chain to go out of bounds, get rejected, and make a copy of its previous step in the process. We make these boundary conditions intuitive and fast in our \chisq \ routines by checking them first and returning an infinite \chisq \ if it is out of bounds. The MCMC routine then automatically assigns a zero likelihood to this step and will always reject it, making a copy of the previous step in the process and preserving the meaning of boundaries.

\begin{figure}[h]
  \begin{center}
    \includegraphics[width=3.25in]{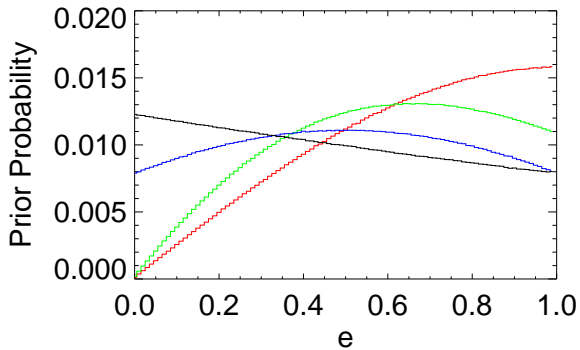} 
    \caption{The prior distribution of eccentricities for each parameterization, when the incorrect implementation of the Metropolis-Hastings algorithm is used, as described in the text, with the same legend as in Figure~\ref{fig:eprior}. The detailed effect this error has depends on the particular parameterization of eccentricity, and the step size, which is ideally equal to the uncertainty. In this plot, the attenuation region is exaggerated because the differential evolution code picks a step size that is large enough to efficiently fill all of the allowed likelihood space, and thus is attenuated by both boundaries at once. However, this attenuation is significant as long as the median value is within a few stepsizes of a boundary. This shows a strong a priori bias against boundaries like $e=0$ when the MH algorithm is improperly implemented.}
    \label{fig:epriorzoom}
  \end{center}
\end{figure}

To demonstrate why copying this step is necessary, we repeat the exercise from \S\ref{sec:eprior}, but without copying the previous step when a step is rejected. The resultant prior distributions of eccentricity are shown in Figure~\ref{fig:epriorzoom}, which shows the prior probability is strongly attenuated near imposed boundaries. This, in turn, will bias the inferred values away from said boundaries. The depth of this attenuated region is proportional to the step size, which is typically equal to the uncertainty in the parameter. We note that the affected region plotted here is exaggerated because the DE-MC code automatically chooses a large stepsize to efficiently fill all of the likelihood space. However, for values near ($\lesssim 3$ times the step size) a boundary, the prior probabilities are still significantly impacted by this bias.

Unfortunately, it is difficult to say how prevalent these latter two biases may or may not be in the literature, or, in the case of the improper MH algorithm, even how significant it is if we know it is present, since the bias depends on the parameterization and even the step size. The larger the step size, the larger the bias (due to the larger attenuated prior region). While the ideal step size is roughly equal to the one-sigma uncertainty, there is no guarantee that a suboptimal step size was not used, because it is not supposed to matter. Nevertheless, some combination of these latter two biases may at least partly explain the large number of nominally-significant eccentricities, even after accounting for the Lucy-Sweeney bias, for systems that are expected to be tidally circularized.
\section{Transit Fitting}
\label{sec:transit}

\subsection{Transit Model}
A primary transit occurs when a planet passes in front of its star and blocks a portion of its light for a period of time. We monitor the star's brightness during this time by taking repeated exposures, and then comparing the target star's brightness to an ensemble of comparison stars. If the star were uniformly bright, the relative flux we would see during transit would be:

\begin{equation}
  F(t) = F_0\left[1 - \lambda^{e}\right]
\end{equation}

\noindent where \fo \ is the baseline flux and analytic equations for $\lambda^{e}$ are defined in \citet[Eq. 1,][]{mandel02}. $\lambda^{e}$ is solely a function of the transit geometry, $\rprs \equiv p$, which is the radius of the planet in stellar radii and $r/R_* \equiv z$, which is the projected distance from the center of the planet to the center of the star, in stellar radii and is a function of time.

\subsection{Limb Darkening}
\label{sec:ld}

In reality, stars are not uniformly bright -- for typical broad-band optical/NIR filters, their apparent brightness falls toward the limb of the star, which is an effect called limb darkening. For main sequence stars, the intensity of the star, $I$, is well-described as functions of $\mu = \cost$, where $\theta_*$ is the angle between the observer and the normal vector on the surface of the star -- that is, $\mu = 1$ at the center of the star, where the normal vector points directly at the observer, and $\mu = 0$ at the limb of the star, where the normal vector is perpendicular to the observer's line of sight.

There are many different ways to parameterize the intensity of the star. We will discuss the most commonly-used laws for transiting planets. The linear limb darkening law,

\begin{equation}
  \label{eq:linearld}
  \frac{I(\mu)}{I(1)} = 1 - u_0(1-\mu),
\end{equation}

\noindent where $u_0$ is a limb-darkening coefficient, was the first obvious choice, but it quickly became clear it was insufficient to describe real surface brightness profiles \citep[e.g.][]{klinglesmith70}. For the precision of many ground based transits, the linear law is still sufficient, but the light curves from {\it Hubble Space Telescope} for HD 209458b showed the linear limb-darkening law to be inadequate for high-precision transit light curves \citep{brown01}. Thus, many have adopted a quadratic limb darkening law of the form:

\begin{equation}
  \label{eq:quadld}
  \frac{I(\mu)}{I(1)} = 1 - u_1(1-\mu) - u_2(1-\mu)^2.
\end{equation}

\citet{mandel02} state that the quadratic limb darkening law is sufficient to describe transit light curves with a precision of $10^{-4}(p/0.1)^2$ -- a precision that has never been achieved from the ground. This was confirmed empirically by \citet{southworth08}, who showed that the quadratic limb darkening law was sufficient for the quality of data then achievable. However, the accuracy of the quadratic law is worse than the 20 parts per million that is achieved routinely from \kepler \ \citep{koch10b} for large planets ($p > 0.04$). For these planets, \kepler \ must therefore use a non-linear limb darkening law of the form,

\begin{equation}
  \label{eq:nlld}
  \frac{I(\mu)}{I(1)} = 1 - a_1(1-\mu^{1/2}) -  a_2(1-\mu) - a_3(1-\mu^{3/2})-  a_4(1-\mu^2).
\end{equation}

Unfortunately, our theoretical predictions of the coefficients has proven insufficient for very precise light curves \citep[e.g][]{knutson07}, and either the limb-darkening needs to be fit by the data, or perhaps newer, more precise models based on 3D hydrodynamical models of stellar atmospheres may be sufficient \citep{hayek12}. Since the transit flux given by the quadratic limb-darkening law is applicable to all but the most precise light curves and is significantly faster to compute than the non-linearly limb-darkened transit flux, we limit our discussion to the quadratic limb darkening law. Note that in the discussion that follows, we can reproduce the linear law precisely by fixing $u_2$ to be zero.

\citet{mandel02} give the quadratically limb-darkened flux during transit as:

\begin{equation}
  \label{eq:transitflux}
  F(t) = F_0\left(1 - \frac{(1-u_1-2u_2)\lambda^{e} + (u_1+2u_2)\left[\lambda^{d} + \frac{2}{3} \Theta(p-z)\right] - u_2\eta^{d}}{1-u_1/3-u_2/6}\right)
\end{equation}

\noindent where $\lambda^{d}$, and $\eta^{d}$ are given in table 1 of \citet{mandel02} for all possible geometries. Like $\lambda^{e}$, $\lambda^{d}$ and $\eta^{d}$ only depend on $p$ and $z$. $\Theta$ is a step function equal to 1 where $p > z$ and 0 elsewhere.

As a side note, in Appendix~\ref{sec:analyticld}, we show that both the quadratic and linear limb darkening flux during transit can be can be written as linear combinations of analytic functions. Therefore, the limb darkening coefficients can be solved analytically for fixed $p$ and $z$.
\subsection{Planetary Path}

Given these analytic expressions, generating a model lightcurve becomes a matter of computing $z$ for all times, which is similar to computing the RV\@. First, we calculate the true anomaly in the same way as before (i.e., using Equations~\ref{eq:trueanom},~\ref{eq:eccenanom}, and~\ref{eq:meananom}). Then, the three-space coordinates of the planet's position relative to the star, as seen from Earth, are

\begin{equation}
  \label{eq:xy}
  \begin{split}
    r =& \frac{a}{R_*}\frac{(1-e^2)}{1+e\cos{\theta(t)}} \\
    X =& -r\cos{\left(\theta(t) + \omega_*\right)} \\
    Y =& -r\sin{\left(\theta(t) + \omega_*\right)\cosi} \\
    Z =&  r\sin{\left(\theta(t) + \omega_*\right)\sini}
  \end{split}
\end{equation}

\noindent where $r$ is the distance from the center of the star to the center of the planet as a function of time, $a$ is the semi-major axis, and \rs \ is the stellar radius. Some have opted to mix $\omega_*$ and $\omega_P$ at this point -- while using $\omega_P$ makes more intuitive sense, we feel the consistent use of one value for the argument of periastron reduces the chance of accidentally misapplying one or the other, and $\omega_*$ is already widely in use.

$Z$ is along the line of sight, where $+Z$ is toward the observer, and the X-Y plane is the plane of the sky. Neither transits nor RV can constrain the longitude of the ascending node, $\Omega$, which is the angle from North to the ascending node, measured counterclockwise (i.e., the rotation of the X-Y plane), but for completeness, the orientation with respect to an observer on Earth is

\begin{equation}
  \label{eq:xyzp}
  \begin{split}
    X' =& -X\cos\Omega + Y\sin\Omega \\
    Y' =& -X\sin\Omega - Y\cos\Omega \\
    Z' =& Z
  \end{split}
\end{equation}

\noindent where $-X'$ is East and $+Y'$ is North. For concreteness, we assume $\Omega=180^\circ$, which means $X=X'$, $Y=Y'$, and $Z=Z'$. This implies that, during primary transit, the planet moves from $-X$ to $+X$ and at $X=0$, $Y$ is equal to the opposite of the impact parameter, $-b$. Finally,

\begin{equation}
  \label{eq:z}
  z = \sqrt{X^2 + Y^2},
\end{equation}

\noindent where $z$ is in units of stellar radii, exactly as required by the \citet{mandel02} code. We must take care to note the sign of $Z$ -- both transits and occultations occur when $z < 1+p$, but it is a primary transit when $Z > 0$, and a secondary eclipse when $Z < 0$.

The calculation of the planetary path is done in our program {\tt EXOFAST\_GETB} and includes the general handling of $\Omega$, which would be useful if one would like to include astrometric measurements.

While we do not support fitting the secondary eclipse, the calculation of its model flux is identical save a couple minor substitutions: $p$ becomes $1/p$, $z$ becomes $z/p$, and the limb darkening of the planet can be ignored.\footnote{This does, however, require the bug fixes described in Appendix~\ref{sec:bugfix}} The resultant model is the observed flux from the planet as it is blocked by the star. In general, this will be some combination of thermal and reflected light, but without knowing the temperature, albedo, and thermal redistribution (e.g., from a phase curve), the two cannot be distinguished, and thus only one additional parameter for each observed bandpass is required for the normalization of the planetary flux (i.e., the eclipse depth). This normalization is a linear parameter, but note the warning above about hybrid linear and non-linear fits. We stress that this is not the same as fitting different values of $p$ for the primary transit and secondary eclipse. The shape of the ingress/egress and the duration of the eclipse require $p$ to be the ratio of radii, not just the square root of the depth.
\subsection{Parameterization}
\label{sec:transitinputs}

We also need to define the parameterization of the transit light curve, which is much less obvious and has been done many different ways in the literature. Most, however, have agreed upon \tc, \fo, \logp, and the quadratic limb darkening parameters $u_1$ and $u_2$. While the eccentricity is also required to derive the model, it is often fixed at the best-fit values from the RV\@. The remaining parameters, which determine the shape of the transit, have no universally-accepted parameterization, likely because each parameterization has its own advantages and disadvantages.

\citet{seager03} suggested $T_T$, the total duration of the transit (first to fourth contact), $T_F$, the duration of the flat part of the transit (second to third contact), and the transit depth, $\delta$ (for non-grazing transits with no limb-darkening, $\delta=p^2$). \citet{carter08} suggested that a less-correlated parameterization would be $\tau$, the duration of ingress or egress (i.e., from first to second contact or third to fourth contact), $T$, the duration from mid-ingress to mid-egress, and $\delta$.

Unfortunately, both of those parameterizations are undefined for grazing geometries, and therefore it is impossible to correct the prior distributions to be physical for all geometries. Grazing transits would be poorly fit, and near-grazing geometries may be unfairly biased by the parameterization (See \citet{carter08} for a discussion). For this reason, we advocate a more physically-motivated parameterization: $\log{(a/R_*)}$, \cosi, and $p$. The advantage to this parameterization is that it intuitively imposes reasonable priors on the physical parameters, and is well-defined for all geometries. The disadvantage is that they are further removed from what is actually measured (the shape of the transit) and the covariances between these parameters is large. Fortunately, the DE-MC algorithm automatically takes the covariances into account, so this is much less important.
\subsection{Other biases}
\label{sec:otherbiases}

A Lucy-Sweeney-like bias exists for all bounded parameters, which includes $\cos{i}$ (in the case of transiting planets when we cannot distinguish between $\pm\cos{i}$), and $p$. Unfortunately, the Lucy-Sweeney-type bias for $\cos{i}$ is unavoidable, but unlike eccentricity, where we expect Hot Jupiters to be tidally circularized and therefore $e$ to be exactly 0, there is no reason to expect a planet to be exactly edge on. Therefore, we are significantly less likely to encounter such a bias. Further, the theoretical interpretation of such a system does not qualitatively change if we measure a small inclination, whereas a small non-zero eccentricity for a planet that is supposed to be tidally circularized requires exotic explanations, such as anomalous values of the tidal Q factor or additional bodies perturbing the system. However, we still need to be careful not to overinterpret nominally-significant impact parameter changes if they are close to 0.

We can avoid a bias in $p$ altogether by thinking about how our model would behave if negative values were allowed. While unphysical, we can make the identification that a negative planetary radius would add flux during transit. To that end, we allow negative planetary radii, calculate the flux decrement as if it were positive, and then add the flux to the baseline rather than subtract it. This avoids the Lucy-Sweeney-type bias, since a negative value of $p$ implies a unique, well-defined likelihood. If the median $p$ is negative with low significance, it is likely there is no transit at all. If it is negative with high significance, there are likely large systematics in the data. If, however, we see a small tail at negative values but the result is statistically significant, we can be more confident that the non-zero measurement is real, and not a result of a bias in fitting. This will be particularly useful when measuring small transits with low significance, whose depths would otherwise tend to be overestimated, just like eccentricity. We could achieve a similar effect for secondary eclipses by allowing the normalization to be negative.

While we choose to step in $p$, the same trick could be played with $\delta$. However, since $p$ is required to calculate the model transit, and $p = \sqrt{\delta}$, we would have to redefine $p=sign(\delta)\sqrt{|\delta|}$ to avoid an imaginary $p$. It is not yet clear which would impose a more physical prior.

One may consider stepping in $\log{p}$, which has no such positive-definite requirement, similar to our steps in $\log{K}$, $\log{P}$, or $\log{a/R_*}$. However, in the case of $p$, systematics in the data may be present which mimic a negative $p$, which is not the case for the other parameters. If our model is not allowed to fit such a systematic, we would unfairly bias the result toward positive values.
\subsection{Transit Best Fit}

Transits are first identified in transit surveys using data from relatively small telescopes that monitor large areas of the sky at once (TrES, \citealt{alonso04}; HATNet, \citealt{bakos04}; XO, \citealt{mccullough06}; CoRoT, \citealt{baglin06}; SuperWASP,
\citealt{cameron07}; KELT, \citealt{pepper07}; Kepler, \citealt{borucki10c}; QES, \citealt{alsubai11}). They use a Box Least-Squares (BLS) algorithm \citep{kovacs02}, which extracts the duration, depth, period, and \tc \ of the transits. For the high-precision transits which this code was designed for, these quantities will be already roughly known, either from the literature or BLS fits to their own survey data. With these parameters, the problem is greatly simplified to finding a local minimum around relatively well-behaved region of parameter space.

With good starting values for $P$ and \tc, we can begin with fairly generic guesses for the rest of the parameters and standard local minimization routines like {\tt AMOEBA} work well to find the local minimum. Once found, we follow the same procedure as the RV data (\S\ref{sec:rv}) and scale the uncertainties to get $\pchisq = 0.5$.

Usually, transit fits of survey-quality data are too degenerate to robustly fit for the impact parameter, and it must instead be fixed to zero (i.e., central crossing). However, because we simultaneously fit the stellar properties (see \S\ref{sec:rvtran}), our code has been tested on KELT survey data and works well when given a good starting value for \tc \ and $P$ from an independent run of BLS\@. Therefore, EXOFAST may also be a useful tool for vetting grazing eclipsing binaries from survey data.
\section{Radial Velocity and Transit}
\label{sec:rvtran}

For simultaneous fits to RV and transit data, the models themselves are the same, but the advantage of fitting the data simultaneously is that they both constrain many of the same parameters, which improves the quality of both fits and ultimately gives us a clearer picture of the system as a whole. Further, we can include additional effects with no penalty, such as the light travel time difference due to the reflex motion of the star (\S\ref{sec:reflex}). More important, covariances between parameters in the different data sets may be unintuitive and non-negligible. Fitting the two data sets separately assumes the covariances between the parameters in the two data sets are zero, whereas a simultaneous fit naturally takes these covariances into account.

\subsection{Parameterization}

The disadvantage of a simultaneous fit is having to rethink the parameterization of the problem, since the overlapping constraints are not always intuitive. The parameters \logp, \secosw, \sesinw, and \tc \ trivially overlap between the two data sets. \logk, $\gamma$, \dvdt, are still mostly independent parameters for RV, and $F_0$, $\cos{i}$, $\log{a/\rs}$, and $p$ are mostly independent for photometry.

However, the combined parameterization is actually a one-parameter family of solutions, meaning that with an estimate of the mass of the star, radius of the star, or a clever combination of the two, we can solve the entire system precisely, including the stellar mass and radius. Of course, the minimum mass of the planet, $\mpl\sin{i}$, cannot be determined from RV without estimating the mass of the primary (even if we assume $\ms \gg \mpl$), and all of the physical parameters from the transit scale with $R_*$ (and we must assume $\ms \gg \mpl$), which the transit cannot constrain \citep{seager03}. Since we must use external information anyway, it behooves us to do it during each step in the Markov chain and use all of the information of the two data sets to their full advantage while simultaneously exploring the covariances between all parameters.

The stellar surface gravity, $g_*$ (often measured as \logg), is equal to  $G\ms/\rs^2$, and is one clever combination of stellar parameters that allows us to break the degeneracy. With that, Kepler's law, and equation~\ref{eq:KP}, the semi-major axis of the planet's orbit, true mass of the planet, mass of the star, and radius of the star, all in physical units with no approximation, become a function of observed quantities:

\begin{equation}
\label{eq:mrstar}
\begin{split}
a &= \frac{g_{*}P^2}{4\pi^2}\left(\frac{\rs}{a}\right)^2 + \frac{KP\sqrt{1-e^2}}{2\pi \sin{i}}, \\
\mpl &= \frac{2\pi Ka^2\sqrt{1-e^2}}{GP\sin{i}}, \\
\ms &= \frac{4\pi^2 a^3}{GP^2} - \mpl, \\
\rs &= a\frac{\rs}{a}. 
\end{split}
\end{equation}

With the approximation that $\ms \sim \ms + \mpl$, the latter terms in the equations for $a$ and \ms \ drop out, and $K$ (i.e., radial velocity) is no longer required, meaning we could apply this technique generally to transit fits alone (of course, losing our constraint on \mpl). In fact, this is how we actually fit transit-only data sets.

However, we have no constraint on \logg \ from the transit or radial velocity alone, so simply adding this parameter to the model without additional information will force the Markov Chain to inefficiently explore this degeneracy. This problem is most usually solved by iterating between light curve and RV fitting and isochrone modeling \citep[e.g.,][]{yi01} to get the model parameters \citep[e.g.,][]{bakos12}, but that sort of iteration does not properly account for covariances between the stellar and planetary parameters. Worse, those fitting follow-up light curves almost always assume the fixed stellar parameters derived from the discovery paper and simply ignore the inconsistency between the stellar density implied by their new light curve and their assumed stellar parameters.

Recently, \citet{torres10} determined an empirical polynomial relation between the masses and radii of stars, and their \logg, effective temperatures, \teff, and metallicities, \feh \ (see their Table 4) based on a large sample non-interacting binary stars in which all of these parameters were well-measured. This is essentially a computationally-convenient way of modeling isochrones, which imposes the same mass-radius constraint to break the degeneracy, but is fast enough to incorporate at each step in the Markov chain. Therefore, we add \logg, \teff, and \feh \ to our stepping parameters and, at each step, we use equation~\ref{eq:mrstar} to derive the self-consistent \ms and \rs that is used to generate the model. Finally, we calculate what the Torres relations would predict for \ms \ and \rs, and apply a prior penalty to the $\chi^2$ for the difference between the Torres values and our model values, using the scatter about their fitted relations ($\sigma_{\log{\ms}} = 0.027$ and $\sigma_{\log{\rs}} = 0.014$), as the prior width.

The constraint on \teff \ and \feh from the Transit data, RV data, and Torres relation is very poor, resulting in highly uncertain values for \ms \ and \rs. Fortunately, \logg, \teff \ and \feh \ can be easily measured with a high-quality spectrum, and can then be applied as priors during each step of the fit for a precise estimate of the stellar parameters. Therefore, our total \chisq \ at each step is

\begin{equation}
\label{eq:priors}
\begin{split}
 \chisq &= \sum_{i=1}^{NRV}\left(\frac{RV_i - Model_{RV,i}}{\sigma_{RV,i}}\right)^2 \\ 
 &+ \sum_{i=1}^{NTransit}\left(\frac{Transit_i - Model_{Transit,i}}{\sigma_{RV,i}}\right)^2 \\ 
 &+ \left(\frac{\ms - M_{Torres}\left(\logg,\teff,\feh\right)}{\sigma_{M,Torres}}\right)^2 \\
 &+ \left(\frac{\rs- R_{Torres}\left(\logg,\teff,\feh\right)}{\sigma_{R,Torres}}\right)^2 \\
 &+ \left(\frac{\logg - \logg_{spec}}{\sigma_{\logg,spec}}\right)^2 + \left(\frac{\teff - T_{spec}}{\sigma_{T,spec}}\right)^2 \\
 &+ \left(\frac{\feh - \feh_{spec}}{\sigma_{\feh,spec}}\right)^2,
\end{split}
\end{equation}

\noindent plus penalties for any other priors we choose to impose. When done this way, the Torres relation, \teff, and \feh, plus the density of the star from the transit can often constrain \logg \ better than its spectroscopic counterpart, more directly and more precisely constraining \ms \ and \rs. These constraints, in turn, feed back directly to the fundamental planetary parameters we care about most (e.g., \mpl \ and \rp). Sometimes, \logg \ can actually be better-constrained by the spectroscopy, in which case that constraint feeds back into the constraint on $a/R_*$, and the other transit parameters.

Adding the ``prior'' penalty to \ms \ and \rs \ from the Torres relation in this manner is somewhat unconventional. Typically, priors are static and come from previously-fit data, not model-dependent and derived at each step. To avoid this, we considered stepping in \logg, \teff, \ and \feh and use the \ms \ and \rs \ from the Torres relation to break the degeneracy, but then we would be solving a one-parameter degeneracy with two parameters, over-constraining the model and leading to inconsistencies. Said another way, the Torres relation is not mathematically self-consistent: the input \logg \ does not precisely equal the \logg \ derived from the output \ms \ and \rs, and therefore there would be multiple ways to calculate critical parameters. Additionally, the theoretical scatter in the Torres relation would set a floor to how well we could measure \ms \ and \rs, regardless of other constraints. Finally, if we were to do it this way, \teff \ and \feh \ would define \ms \ and \rs, not simply constrain it. Therefore, we would lose the power to influence \teff \ and \feh away from their prior values. While typically, the parameters derived in different ways are statistically consistent, as we would expect, if we use the output \rs \ and \ms \ to generate any piece of our model, rather than as a prior constraint, the model itself would be mathematically inconsistent. Applying these model-dependent priors uses the information encoded in the Torres relation, which may more appropriately be thought of as data which we merely expect to be statistically consistent, while maintaining the mathematical self-consistency of the model.

\citet{enoch10} recasts the Torres relation in terms of \rhostar \ instead of \logg, with the idea being that the transit precisely constrains \rhostar \ (ignoring \mpl), and so it is a more efficient stepping parameter for the Markov chain, which is true. Unfortunately, for the same reason the transit precisely constrains \rhostar, we can no longer derive independent constraints on \ms \ and \rs. Therefore, we would be required to use the output \ms \ and \rs \ to link the RV and transit models, and we are left with a mathematically inconsistent model (their input \rhostar \ does not equal the \rhostar \ derived from the output \ms \ and \rs \ either), and we are left with no additional constraint on \teff \ and \feh.

One reasonable criticism of our method is that it assumes the Torres mass and radius relations are independent. To the extent this is not true, we are double-counting the prior imposed from the Torres relations, and artificially reducing our errors. Alternatively, we can discard a relation, either for \rs \ or for \ms, but then we assume the relations are perfectly correlated and we lose information to the extent that they are not. A more correct method likely lies somewhere in between, explicitly accounting for the covariances between the \ms \ and \rs \ relations.

A consequence of using the Torres relations (or any stellar model) is that we inherit their limitations -- i.e., EXOFAST, in its current form, should only be applied to ``single (post-) main-sequence stars above 0.6 M$_\sun$'' \citep{torres10}. For stars outside of this regime, the prior for the Torres relation should be removed, and an independent prior on the stellar mass and/or radius should be imposed by editing the \chisq \ function (a trivial task assuming another measurement is available). The code will issue a warning if this regime is encountered, which can be safely ignored if it is only issued during the burn-in period.

A final note about this procedure is that the covariance between \logg \ and $a/R_*$ is extreme because they only differ by one power of \rs. If this covariance is not accounted for when stepping in the Markov Chain (e.g., with the DE-MC algorithm), the Markov Chain becomes extraordinarily inefficient, taking hundreds of times longer than it otherwise would.
\subsection{Limb Darkening}

With the values of \logg, \teff, \ and \feh \ from the steps in the Markov chain (and the observed bandpass), we can also derive the limb-darkening parameters by interpolating the tables from \citet{claret11}. This interpolation is done in our module {\tt QUADLD}. However, the tables contain a substantial and poorly-quantified systematic error. If this error were not taken into account, we would underestimate the errors on any covariant parameters. Worse, if the data were of sufficient quality to constrain the limb-darkening parameters, they would become an implicit constraint on \logg, \teff, and \feh \ (and therefore the mass and radius of the star). Since we know the limb darkening tables to be flawed \citep{hayek12}, this would in turn bias the stellar parameters and all derived parameters. To account for this, we do something similar to what we described above with the Torres relation. We estimate the error on the quadratic limb darkening parameters, $\sigma_{u,1}$ and $\sigma_{u,2}$ to be 0.05, based on Figure 1 of \citet{claret11}. Then, we step in both limb darkening parameters, $u_1$ and $u_2$,  calculate what the \citet{claret11} tables would predict from the current steps in \logg, \teff, and \feh, for the limb darkening coefficients, $u_{1,Claret}$ and $u_{2,Claret}$, and add an additional penalty to the \chisq \ (Equation~\ref{eq:priors}) of the form:

\begin{equation}
\label{eq:ldchi2}
\left(\frac{u_1 - u_{1,Claret}}{\sigma_{u,1}}\right)^2 + \left(\frac{u_2 - u_{2,Claret}}{\sigma_{u,2}}\right)^2.
\end{equation}

Ideally, we would like to have accurate errors for the limb darkening coefficients and use more accurate, non-linear limb darkening tables from 3D hydrodynamical models, but neither are currently available. Additionally, a non-linear limb darkening model is currently impractically slow to calculate for large data sets, as our speed enhancements described in Appendix~\ref{sec:occultquad} would no longer apply.

If an accurate grid of theoretical limb darkening parameters with well-understood uncertainties were available for a wide range of stars, EXOFAST may be able to work backwards, and use the limb darkening to constrain \logg, \teff, and \feh -- and therefore the masses and radii of stars -- from a precise light curve alone -- a tantalizing possibility. In Appendix~\ref{sec:analyticld}, we describe how one could linearly fit the quadratic limb-darkening parameters, which may prove useful with such a scheme.
\subsection{Reflex Motion}
\label{sec:reflex}

Having the semi-major axis in physical units provides an absolute scale to the system, which allows us to include the reflex motion of the star without introducing any additional parameters. Equation~\ref{eq:xy} assumes a barycentric coordinate system, but the RVs are measured in the stellar-centric frame, and the transits are measured in the planet-centric frame. The light travel time (Roemer delay) between these frames can be large, and is analagous to the standard correction from the Heliocentric Julian Date (HJD) or Julian Date (JD) to Barycentric Julian Date (BJD) in our own solar system \citep{eastman10b}.

One can attempt to correct the flux of the transit \citep{loeb05} or the observed radial velocity to account for this,\footnote{We could even transform the model flux or model radial velocity to truly as-measured, geocentric frame!} but it is far more straight-forward to transform the observed time stamp into the target-barycentric frame, which is done at each step in the Markov Chain with our code {\tt BJD2TARGET}.

This effect is most important when trying to reconcile observations for which the information comes from different points in space, such as the primary transit, secondary eclipse, radial velocity, phase curves, or multiple transits of different planets around the same star. For a typical Hot Jupiter ($a\sim0.05$ AU) the difference between different reference frames is $\sim30$ seconds, and it is obviously much larger for planets farther from their stars.

Even in the case of a single observation of a transiting Hot Jupiter, the duration of a transit (as seen in the Solar System Barycenter frame) differs by 0.15 seconds from its true value if this effect is ignored, which is only marginally negligible by today's standards. For Hot Jupiters, the offset between the stellar-centric and barycentric frame is negligible ($\sim$10 ms), but grows with the semi-major axis and the mass of the companion.

While the target-barycentric times are important for dynamical studies of Transit Timing Variations (TTVs), they are really only necessary when there are two transiting planets, which our code does not consider. With a single planet, the offset between the frames for all primary transits will be constant, so the TTVs will not be different. To avoid confusion, and because the times in the Solar System Barycentric frame (i.e., \bjdtdb) are  most precisely known\footnote{i.e., it does not include the uncertainty the light travel time due to the uncertainty in the target's semi-major axis}, we only quote the times in \bjdtdb.

This conversion back to \bjdtdb \ is done by our module, {\tt TARGET2BJD}. To be clear, this procedure automatically reconciles the locations of each observation, which for example, naturally constrains the eccentricity from the primary transit and secondary eclipse timing.

We could calculate the Roemer delay due to the systemic velocity, $\gamma$, with no additional parameters. Since the system may be receding or approaching us, the systemic velocity will tend to stretch out or compress the arrival time of the photons by a constant factor, $1 + \gamma/c$. However, since RVs are often arbitrarily normalized, we do not always have a good measurement of the absolute systemic velocity. More important, the only practical effect such a correction would be to change the period by a factor of $1+\gamma/c$. While this $\lesssim0.1\%$ difference in period is already highly statistically significant in many systems, the practical implications of such a difference will be completely obscured by the uncertainties in the other parameters for the foreseeable future, and quoting this corrected period would only make deriving the ephemeris in the observed frame more error-prone. Therefore, we actively chose not to correct for the additional Roemer delay introduced by $\gamma$.

There are several other minor effects on the arrival time of the photons that we ignore, because they would require additional free parameters which are generally not well-known and the effect is very small. These include the light travel time from our barycentric frame to the target barycentric frame, proper motion \citep{rafikov09}, parallax \citep{scharf07}, and general relativistic precession \citep{jordan08, pal08a}.

\subsection{Ignored Effects}

The ignored effects are numerous, and include the Rossiter-Mclaughlin (RM) effect, Transit Timing Variations (TTVs) (if multiple epochs are fit simultaneously, a constant ephemeris is assumed), secondary eclipses, a distance constraint on the stellar radius, reflection from the planet, ellipsoidal variations of the star, relativistic beaming, lensing, gravity brightening, non-Keplerian gravitational interactions, and tidal effects, to name a few. It is expected that for a given system, many will want to modify the code slightly to include some of these additional effects. We have attempted to make the code as modular and comprehensible as possible, such that someone familiar with IDL could use EXOFAST as a starting point or template for their own, more specialized code without necessarily needing to master the details of modeling. Future versions of our code may include some of these ignored effects as we have occasion to worry about them, and those codes may be made publicly available.

\subsection{Combined Fit}
\label{sec:combined}

Now, with a mere 16 parameters, $\gamma$, \dvdt, \tc, \logp, $\sqrt{e}\cos{\omega_*}$, $\sqrt{e}\sin{\omega_*}$, $\log{K}$, \cosi, $p$, \fo, $\log{a/R_*}$, \logg, \teff, \feh, $u_1$, and $u_2$, we can describe the both the radial velocity and transit data simultaneously, including often neglected effects.

Since we have a very good guess for the starting values for each parameter from the individual fits, no global fit is required. Assuming they are consistent with one another, we only need a slight refinement to find the best fit of the combined data sets. Therefore, using the starting values for the best fits we found in both individual cases, we run a quick {\tt AMOEBA} minimization to find the local minimum of the combined data set, and we are finally ready to use our MCMC code, {\tt EXOFAST\_DEMC}, to determine the model uncertainties and covariances.

To summarize, our procedure to calculate the \chisq, after we have stepped in the 16 parameters above, is as follows:

\begin{enumerate}
\item Check the boundary conditions for each parameter. If any parameter is out of bounds, we immediately return $\chisq =\infty$. 
\item Compute $a$, \rs, \mpl, and \ms \ (Equation~\ref{eq:mrstar}).
\item Compute \ms \ and \rs \ from the Torres relation.
\item Interpolate the \citet{claret11} tables to get the quadratic limb darkening parameters at the given \logg, \teff, and \feh.
\item Compute \tp.
\item Correct the observed times to the target's barycentric frame (\S\ref{sec:reflex}).
\item Calculate the RV model (\S\ref{sec:rv}).
\item Calculate the transit model (\S\ref{sec:transit}).
\item Compute the total \chisq \ (Equations~\ref{eq:priors} and~\ref{eq:ldchi2}).
\end{enumerate}
\section{An Example}
\label{sec:example}

To explain how to use the code, describe its outputs, and validate it at the same time, we now follow an example fit from beginning to end using RVs and transit data of HAT-P-3b from \citet{torres07} (T07 hereafter), hosted on the exoplanet archive. This system was chosen nearly randomly -- we just looked down the list of HAT planets and took the first one with enough public data and only one source for RVs (since the public version of EXOFAST does not fit multiple RV zero points).

Since the transit data were in \bjdutc \ and magnitudes, we converted it to the required format of \bjdtdb, and normalized flux, and wrote them to a file called ``hat3.flux.''' We also converted the times of the RV data points from \hjdutc \ to \bjdtdb \ and wrote them to a file, ``hat3.rv.''

Next, we have to make decisions about what information we have the power to constrain and which parameters should be constrained by priors from external information, or held fixed. As is always the case, we must adopt priors for \teff \ and \feh \ based on spectra, which T07 quote in their Table 2. As we explain later, we also choose to use the spectroscopic constraint on \logg. For all spectroscopic priors, we use the larger uncertainties they describe in the text. Like T07, due to relatively poor phase coverage, there was no power to constrain the eccentricity, so we fixed it to zero and did not attempt to fit a slope. The relatively short range of data for the RV meant the transit survey data constrained the period significantly better, so T07 fixed their period at that value. Instead, we use the value from their photometry as a prior. Note that a suitable fit was possible without this prior; the uncertainties were just larger. Finally, we must specify the band in which the transit was observed, in this case, Sloan i'.

Once we have decided on what to fit, we simply call the code to reflect our wishes:

\smallskip
{\tt
EXOFAST,TRANPATH=`hat3.flux',/NOSLOPE,\$\\
/CIRCULAR,PRIORS=priors,BAND=`Sloani',\$\\
RVPATH=`hat3.rv',PNAME=`HAT-P-3b',\$\\
MINP=2.85,MAXP=2.95
}
\smallskip

\noindent where the name of the observed bandpass is given as ``BAND'', ``PNAME'' is the case-insensitive name of a planet in exoplanets.org, from which the starting values or spectroscopic priors can be pulled. This is generally not necessary, but gives the fit a more robust starting point. The ``PRIORS'' input is a 2D array containing the prior value and 1-sigma uncertainty for each parameter. To specify a parameter with no prior, a width of infinity should be used. The parameters ``MINP'' and ``MAXP'' limit the range of the Lomb-Scargle periodogram, and is typically not necessary, but these RV data are too sparse to get reliable results from the periodogram. Further details on the calling structure and additional features can be found in the documentation of the code.

EXOFAST then fits the RV data as described in \S\ref{sec:rv}, scales the uncertainties, fits the transit data as described in \S\ref{sec:transit}, scales its uncertainties, then fits the two data sets combined. It determines the appropriate stepping scale, and begins the Markov chains, giving a status update about the number of accepted steps, and estimates the time to completion, if it were to take the maximum number of steps. Once it has taken 5\% of the maximum number of allowed steps, it will estimate if the chains will be well-mixed by the time it is done. If it is expected to be done before it hits the limit, it will estimate when. If not, it will recommend a thinning factor. In our case, the chains were well-mixed in about five minutes on a standard desktop computer. This is slightly faster than a general run because the orbit was circular. This eliminates two free parameters and makes the normally expensive solution to Kepler's equation trivial. Still, for a similar number of points, ten minutes is fairly typical, even when eccentricity is included.

The estimated time of completion, or the recommended thinning factor is very rough, and should only be trusted to a factor of 2-5. While the thinning factor of $N_{THIN}$ means it may take up to $N_{THIN}$ times longer, it will stop as soon as it is well mixed, so it pays to be a little conservative -- for that reason, the suggested value is twice what it actually calculates it needs. The mixing criteria described in \S\ref{sec:mcmc} are very conservative. Indeed, we have run chains with fewer than 100 independent steps (as opposed to the required 1000) that did not differ significantly from chains that were fully mixed. Therefore, if you find that the recommended thinning value would imply a prohibitively-long execution time, you may wish to proceed with caution in interpreting a chain that gives such a warning. Otherwise, restarting the chain with the suggested thinning factor is highly recommended. Alternatively, if you have a 64-bit machine (and a 64 bit version of IDL), you could increase the maximum number of steps by this same factor, but there is very little difference in the quality of the final output or the execution time, and thinning makes the chains smaller and more manageable without throwing away much useful information.
\subsection{Outputs}

\begin{figure}[h]
  \begin{center}
    \includegraphics[width=3.25in]{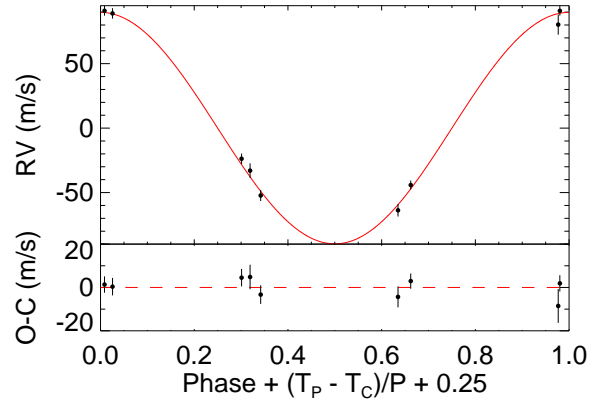} 
    \caption{The best-fit of HAT-P-3 RV data from T07, as output by EXOFAST, forced to a circular orbit and with no slope. The units of the x-axis are chosen so primary transit will always be at 0.25, and for circular orbits (as in this case), secondary eclipse will be at 0.75.}
    \label{fig:hat3rv}
  \end{center}
\end{figure}

\begin{figure}[h]
  \begin{center}
    \includegraphics[width=3.25in]{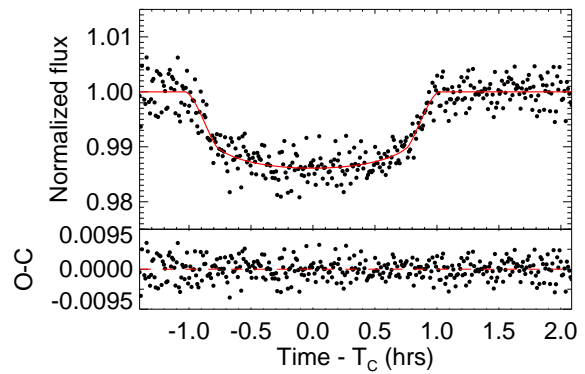}
    \caption{The best-fit transit HAT-P-3b data from T07, as output by EXOFAST\@.}
    \label{fig:hat3transit}
  \end{center}
\end{figure}

Once the program runs to completion, there are several outputs. First, are publication-ready plots of the data with the best-fit model overplotted, and residuals below, as shown in Figure~\ref{fig:hat3rv} and~\ref{fig:hat3transit} for our example.

By setting the debug flag, these plots can be drawn to the screen during each call to the \chisq \ routine. As the name implies, it is useful for debugging if the fit is not working. It can also be instructive -- if, for example, the debug flag is left on during an AMOEBA fit, it will essentially play a movie of the routine settling into the best fit. In this manner, one can gain an intuitive feel for how the algorithm works (or why it is failing). It is also instructive to leave it on at the beginning of an MCMC fit, to gain intuition for the difference between the steady convergence toward the best fit of the AMOEBA algorithm versus the more stochastic MCMC algorithm, which wanders around the vicinity of the best fit. Obviously, this slows down the fit by orders of magnitude and is not intended to be used during a standard fit.

Perhaps the most powerful aspect of MCMC is the fact that, with the parameter chains in hand, it is trivial to generate median values, uncertainties, and covariances for derived quantities -- we simply perform the appropriate transformation to each step in the entire chain.\footnote{Note that our priors are not likely uniform in any of these derived parameters, and we should be aware of that as we interpret these values.} As such, we quote the median values, along with their 68\% confidence intervals for several other derived quantities of interest as a publication-ready LaTeX source code file for the deluxe table shown in Table~\ref{tab:HAT-P-3b}. {\tt EXOFAST\_LATEXTAB}, which generates these tables, automatically rounds the two-sided uncertainties to two significant digits each and rounds the median value to the higher precision uncertainty. When the uncertainty is symmetric, we use the $\pm$ symbol; otherwise, we list both uncertainties in the standard fashion. This fast and accurate way to go from parameter arrays to properly-formatted latex source code is not just convenient, but also mitigates the risk of typos introducing catastrophic errors. Even if it is not intended to be inserted into a paper, it is often more readable as appropriately-rounded LATEX code than simply printing the median values, particularly since it reorganizes the fitted and derived quantities into a more intuitive layout, as we have done in Table~\ref{tab:HAT-P-3b} to group parameters into stellar, planetary, RV, transit, and eclipse categories.
\begin{deluxetable}{lcc}
\tablecaption{Median values and 68\% confidence interval for HAT-P-3b}
\tablehead{\colhead{~~~Parameter} & \colhead{Units} & \colhead{Value}}
\startdata
\sidehead{Stellar Parameters:}
                           ~~~$M_{*}$\dotfill &Mass (\msun)\dotfill & $0.907_{-0.047}^{+0.050}$\\
                         ~~~$R_{*}$\dotfill &Radius (\rsun)\dotfill & $0.772_{-0.041}^{+0.045}$\\
                     ~~~$L_{*}$\dotfill &Luminosity (\lsun)\dotfill & $0.387_{-0.050}^{+0.058}$\\
                         ~~~$\rho_*$\dotfill &Density (cgs)\dotfill & $2.78_{-0.39}^{+0.44}$\\
              ~~~$\log(g_*)$\dotfill &Surface gravity (cgs)\dotfill & $4.620\pm0.042$\\
              ~~~$\teff$\dotfill &Effective temperature (K)\dotfill & $5182\pm79$\\
                              ~~~$\feh$\dotfill &Metalicity\dotfill & $0.271_{-0.079}^{+0.080}$\\
\sidehead{Planetary Parameters:}
                              ~~~$P$\dotfill &Period (days)\dotfill & $2.899703\pm0.000053$\\
                       ~~~$a$\dotfill &Semi-major axis (AU)\dotfill & $0.03852_{-0.00068}^{+0.00070}$\\
                             ~~~$M_{P}$\dotfill &Mass (\mj)\dotfill & $0.591_{-0.024}^{+0.025}$\\
                           ~~~$R_{P}$\dotfill &Radius (\rj)\dotfill & $0.825_{-0.055}^{+0.061}$\\
                       ~~~$\rho_{P}$\dotfill &Density (cgs)\dotfill & $1.30_{-0.24}^{+0.28}$\\
                  ~~~$\log(g_{P})$\dotfill &Surface gravity\dotfill & $3.332_{-0.058}^{+0.056}$\\
           ~~~$T_{eq}$\dotfill &Equilibrium Temperature (K)\dotfill & $1118_{-33}^{+35}$\\
                       ~~~$\Theta$\dotfill &Safronov Number\dotfill & $0.0607_{-0.0043}^{+0.0046}$\\
               ~~~$\fave$\dotfill &Incident flux (\fluxcgs)\dotfill & $0.355_{-0.041}^{+0.046}$\\
\sidehead{RV Parameters:}
                    ~~~$K$\dotfill &RV semi-amplitude (m/s)\dotfill & $89.7\pm1.9$\\
                 ~~~$M_P\sin i$\dotfill &Minimum mass (\mj)\dotfill & $0.590_{-0.024}^{+0.025}$\\
                       ~~~$M_{P}/M_{*}$\dotfill &Mass ratio\dotfill & $0.000622\pm0.000017$\\
               ~~~$\gamma$\dotfill &Systemic velocity (m/s)\dotfill & $-14.3_{-1.4}^{+1.5}$\\
\sidehead{Primary Transit Parameters:}
                ~~~$T_C$\dotfill &Time of transit (\bjdtdb)\dotfill & $2454218.76037\pm0.00033$\\
~~~$R_{P}/R_{*}$\dotfill &Radius of planet in stellar radii\dotfill & $0.1098_{-0.0020}^{+0.0021}$\\
     ~~~$a/R_{*}$\dotfill &Semi-major axis in stellar radii\dotfill & $10.73_{-0.53}^{+0.54}$\\
              ~~~$u_1$\dotfill &linear limb-darkening coeff\dotfill & $0.434\pm0.047$\\
           ~~~$u_2$\dotfill &quadratic limb-darkening coeff\dotfill & $0.225_{-0.049}^{+0.047}$\\
                      ~~~$i$\dotfill &Inclination (degrees)\dotfill & $87.38_{-0.54}^{+0.59}$\\
                           ~~~$b$\dotfill &Impact Parameter\dotfill & $0.490_{-0.092}^{+0.073}$\\
                         ~~~$\delta$\dotfill &Transit depth\dotfill & $0.01206_{-0.00043}^{+0.00046}$\\
                ~~~$T_{FWHM}$\dotfill &FWHM duration (days)\dotfill & $0.07492_{-0.0010}^{+0.00096}$\\
          ~~~$\tau$\dotfill &Ingress/egress duration (days)\dotfill & $0.0109_{-0.0012}^{+0.0014}$\\
                 ~~~$T_{14}$\dotfill &Total duration (days)\dotfill & $0.0859_{-0.0013}^{+0.0014}$\\
      ~~~$P_{T}$\dotfill &A priori non-grazing transit prob\dotfill & $0.0641_{-0.0062}^{+0.0071}$\\
                ~~~$P_{T,G}$\dotfill &A priori transit prob\dotfill & $0.0799_{-0.0079}^{+0.0091}$\\
                            ~~~$F_0$\dotfill &Baseline flux\dotfill & $1.00630\pm0.00019$\\
\sidehead{Secondary Eclipse Parameters:}
              ~~~$T_{S}$\dotfill &Time of eclipse (\bjdtdb)\dotfill & $2454220.21022\pm0.00033$
\enddata
\label{tab:HAT-P-3b}
\end{deluxetable}

\begin{figure*}[h]
  \begin{center}
    \includegraphics[width=6.25in]{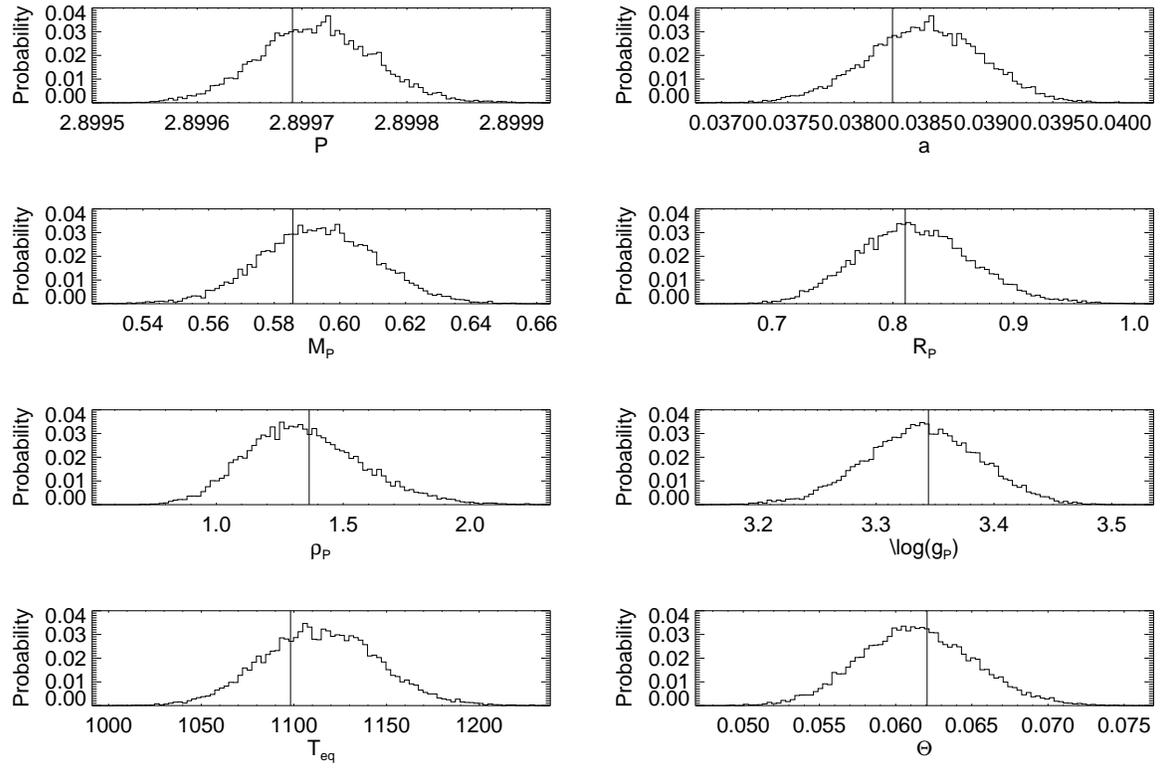}
    \caption{A demonstration of the PDFs output by EXOFAST\@. Normally, these are not intended for publication, but for diagnostics. This shows a subset of the parameter distributions for the combined RV + transit fit to the T07 data of HAT-P-3b, as generated by {\tt EXOFAST\_PLOTDIST}. The line marks the best-fit values corresponding to the minimum \chisq \ amongst all steps in the Markov Chain. The slight roughness in these PDFs would smooth out with longer chains, but the convergence criteria ensure that longer chains are not necessary to determine the median values or their uncertainties reliably.}
    \label{fig:hat3pdf}
  \end{center}
\end{figure*}

However, just because the program will provide an answer that is easy to publish does not necessarily mean it is publication-ready. It is always wise to inspect the parameter distributions for strange behavior that may compromise the results, and check that the parameters themselves make sense. As such, we also output the probability distribution functions for each parameter, including derived parameters -- a subset of which is shown in Figure~\ref{fig:hat3pdf}. The best-fit values (the lowest \chisq \ achieved by the Markov chain) are shown as a solid vertical line over each distribution. These plots are not typically expected to be published, but are rather for diagnostic purposes only.

\begin{figure*}[h]
  \begin{center}
    \includegraphics[width=6.25in]{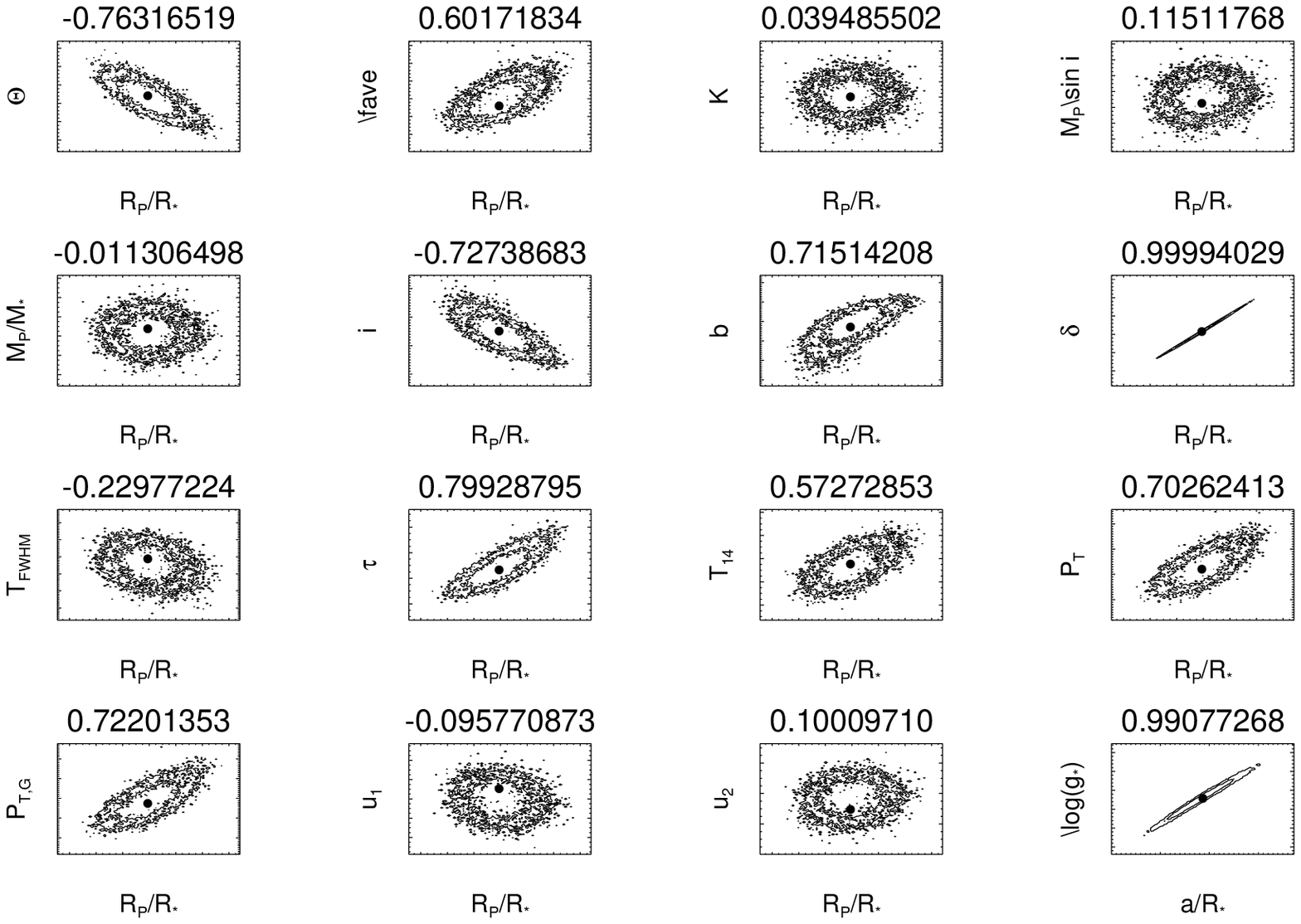} 
   \caption{A demonstration of a small subset of the covariance plots output by EXOFAST\@. Normally, these are not intended for publication, but for diagnostics. This shows some of the covariances for the combined RV + transit fit of HAT-P-3b. The contours are the 68\% and 95\% probability contours, and the black dot is the best-fit value. The number above each plot is the covariance between the parameters.}
    \label{fig:hat3covar}
  \end{center}
\end{figure*}

Another interesting diagnostic is the plot of covariances, a subset of which are shown in Figure~\ref{fig:hat3covar}. We plot contour plots of each parameter against each other parameters, including derived parameters, where the contours show the 68\% and 95\% confidence intervals. We hide the numerical values on the axis labels for readability -- the shape is the most important diagnostic here. The value above the plot is the correlation statistic. The solid black dot is the best-fit value of the two parameters. We note that our routine to generate these contours, {\tt EXOFAST\_ERRELL}, is not a standard IDL routine and may be of general interest. Again, the roughness of these covariance plots would smooth out with longer chains.

The parameter chains themselves may be useful for additional diagnostics or analysis not performed by {\tt EXOFAST}, such as creating publication-quality plots of particular parameter distributions or covariances. We output an IDL save file that includes the array of steps populated by the Markov chain, including all derived quantities. This array includes the burn-in, in order to maintain a complete record of all steps. The save file also includes the corresponding \chisq \ at each step, the latex-style names of all parameters, and the index of the first useful link in the chain (before which is considered the ``burn-in'').
\subsection{Validation}
\label{sec:validation}

Generally, our results are in very good agreement with those found by T07. All quoted values agree with theirs within $1 \sigma$, and the vast majority agree to better than $0.25 \sigma$. The determination of the stellar properties is the most fundamental difference between our two methods. We use the relations from \citet{torres10} and the spectroscopic priors, to enforce consistency with our transit and RV data at each step, whereas they fit the transit, then use the fitted density as a constraint to their stellar isochrones \citep{yi01} in a later step to derive the stellar properties and iterate \citep{sozzetti07}. While this is an attempt to do a similar thing, their process results in statistically consistent, but not identical values for \rhostar \ from the stellar parameters ($2.36$ g cm$^{-3}$) and the \rhostar \ from the fitted transit parameters ($2.67$ g cm$^{-3}$). However, our densities from the two sets of parameters are necessarily equivalent ($2.78$ g cm$^{-3}$).

The uncertainty in the transit time from T07 is almost an order of magnitude larger than what we find. Given the good agreement between everything else, we suspect a typo in T07 is to blame (e.g., a missing zero), as our uncertainty is more in line with analytic estimates \citep{carter08} -- thus demonstrating a major benefit of a code that automatically generates the LaTeX source code of the final parameters directly from the data.

Other than that, all of our uncertainties agree within 30\%, with neither of us finding systematically smaller uncertainties across all parameters. Unsurprisingly, the largest differences are for the stellar parameters, for which we used fundamentally different methods. While we in principle, can derive uncertainties that are smaller than the scatter in the \citet{torres10} relation due to the additional constraints of the RV and transit data, our uncertainties are still dominated by that scatter. These propagate to slightly larger uncertainties in the  planetary parameters.

The radial velocity portion of this code has been used to fit the radial velocity data in \citet{fleming10}, \citet{lee11}, and \citet{wisniewski12}, though minor modifications were required to fit different zero points for each instrument, as described in the text of those papers.

We used this code to fit the KELT-1b data \citep{siverd12}, adding the ability to fit an arbitrary number of transits (with TTVs), RV zero points, and the Rossiter Mclaughlin effect. We also modified our code to fit the KELT-2Ab and KELT-3b \citep{beatty12, pepper12} with an arbitrary number of transits, deblend the stars, and constrain the stellar radius through the {\it Hipparcos} distance prior.

As described in \S\ref{sec:nsteps}, we ran 100 fits of simulated RV curves for each of 15 different intrinsic eccentricities (1500 total fits in our preferred parameterization). With the exception of the Lucy-Sweeney bias for eccentricity discussed in detail there, all measured values for the other 5 parameters were statistically consistent with the input values, finding only 312 of the 7500 parameters outside of $2\sigma$ (341 expected), and 10 outside of $3\sigma$ (20 expected). Further, all of these simulations were robustly fit without any intervention or assumption of the input values.
\section{Online Tools}
\label{sec:online}
\subsection{READEXO}

All of the following routines make use of {\tt READEXO}, an IDL program to read the current exoplanets.org database \citep{wright11} into an IDL structure. It is available for download and use offline as well. Each header entry, described on their website\footnote{http://exoplanets.org/help/common/data}, becomes a unique tag name for the structure. The code will automatically adapt to the addition of rows and columns into the exoplanets.org database, and a flag can be set to automatically update the local copy. This code is useful for more general manipulation and comparison of all exoplanetary data than is allowed by their web interface, and for integration into software suites. In our code, we use it to seed the fits for known planets to make the fits more robust, and for retrieving priors on \logg, \teff \ and \feh \ to get the host properties, rather than requiring the user to supply them. However, we caution strongly that the selection effects inherent to this sample of planets are very poorly understood and poorly quantified -- care should be taken not to over-interpret results that these tools make trivial. If you use this code, please also cite \citet{wright11} and acknowledge their efforts in making and maintaining this incredibly useful database.
\subsection{Ephemerides}

There are now so many transiting planets that, on any given night from any given location on Earth, we are very likely to be able to observe at least one of them \citep{eastman10a}. It is useful, then, to quickly determine which those are in order to plan observations or fill gaps in observing schedules. We present an online calculator that determines which planets are transiting or eclipsing from a particular observatory on a particular date\footnote{http://astroutils.astronomy.ohio-state.edu/exofast/ephem.shtml}. This tool uses the exoplanets.org database, which is synced daily. For the predicted secondary eclipse times of non-circular orbits and especially the transit times of RV planets, the precision is likely to be much worse than what we could derive from the original data because these times were derived from the values of $P$, $e$, $\omega_*$ given in exoplanets.org and thus do not include the often large covariances between these parameters. These could be greatly improved if, in the future, published results included \tc \ and \ts \ directly from the fits, like HAT does \citep[e.g.][]{bakos12}, and these were included in the exoplanets.org database.
\subsection{RV and Transit fitting}

We provide an intuitive online interface to the basic features of EXOFAST\footnote{http://astroutils.astronomy.ohio-state.edu/exofast/exofast.shtml}, which can fit either or both of the RV and transit, including an arbitrary number of systematics. Priors on \teff \ and \feh \ are required, but can be automatically pulled from the exoplanets.org database by selecting the appropriate planet from a pull down menu. This pull down menu will update daily as new planets are added to the exoplanets.org database. When fitted, the transit (normalized to 1 and with systematics removed) and the RV will be plotted and both the fitted and derived parameters will be output to the screen.

Unfortunately, since these are run on the server side, it is not practical to support the full DE-MC code, so only the best-fit values are reported. Therefore, this online fitter is not intended for research-quality fits.
\subsection{Quadratic limb darkening}

Our last online code linearly interpolates the \citet{claret11} quadratic limb darkening tables for given values of \logg, \teff, and \feh \ in any of the following bands: Johnson/Cousins $U$, $B$, $V$, $R$, $I$, $J$, $H$, and $K$; Sloan $u'$, $g'$, $r'$, $i'$, and $z'$; \spitzer \ 3.6 \um, 4.5 \um, 5.8 \um, and 8.0 \um; {\it Kepler}; {\it CoRoT}; and Stroemgren $u$, $v$, $b$, and $y$\footnote{http://astroutils.astronomy.ohio-state.edu/exofast/limbdark.shtml}. This can also draw the stellar parameters from the exoplanets.org database, if available.

\begin{acknowledgements}

We would like to thank Howard Relles, Rachel Ross, Karen Collins, Thomas Beatty, and Trey Mack for beta tests, bug reports, and excellent suggestions for improvements; Eric Ford for CUDA translations of bottlenecked routines, Robert Siverd for useful discussions about the eccentricity parameterization and random number generator; and Wayne Landsman for curating the IDL astronomy library, from which we use many functions \citep{landsman93}. This research has made use of the Exoplanet Orbit Database and the Exoplanet Data Explorer at exoplanets.org. This research has made use of the NASA Exoplanet Archive, which is operated by the California Institute of Technology, under contract with the National Aeronautics and Space Administration under the Exoplanet Exploration Program. This work was supported in part by NSF CAREER grants AST-1056524 and AST-0645416.

\end{acknowledgements}

\appendix

\section{Analytic fit of the linear and quadratic limb darkening}
\label{sec:analyticld}

Given the quadratic limb darkening law (Eq.~\ref{eq:quadld}) and the flux during transit (Eq.~\ref{eq:transitflux}), if we make the following definitions

\begin{equation}
  \begin{split}
    \label{eq:lddef}
    x_0 &= 1-\lambda^{e}, \\
    x_1 &= 2/3(\lambda^{e} - \Theta(p-z)) - \lambda^{d}, \\
    x_2 &= \eta^{d} + \lambda^{e}/2, \\
    c_0 &= F_0, \\
    c_1 &= F_0\frac{u_1 + 2u_2}{1 - u_1/3 - u_2/6}, \\
    c_2 &= F_0\frac{u_2}{1 - u_1/3 - u_2/6},
  \end{split}
\end{equation}

\noindent the flux during transit can be written:

\begin{equation}
  F = c_0x_0 + c_1x_1 + c_2x_2.
\end{equation}

Since $x_0$, $x_1$, and $x_2$, are solely functions of the transit geometry ($p$ and $z$), they can be calculated independent of the limb darkening and are now optional outputs of {\tt EXOFAST\_OCCULTQUAD}. Using these, we can analytically solve for the coefficients $c_0$, $c_1$, and $c_2$ by performing a standard linear \chisq \ minimization \citep[see][]{gould03}. Then the limb darkening coefficients, $u_1$ and $u_2$ simply become:

\begin{equation}
  \begin{split}
    \label{eq:ldcoeffs}
    u_1 = \frac{c_1-2c_2}{c_0 + c_1/3 - c_2/2} \\
    u_2 = \frac{c_2}{c_0 + c_1/3 - c_2/2}
  \end{split}
\end{equation}

This analytic fit can greatly increase the speed of any fitting routine by reducing the dimensionality of the non-linear solver, particularly in cases where data are taken with multiple bands. However, in the low signal-to-noise regime where the data have little power to constrain the limb darkening, this analytic fit can give non-physical results, in which case it is better to fit a linear limb darkening law, fix the values at the \citet{claret11} values, fit them non-linearly with a prior, or allow them to vary by interpolating the \citet{claret11} limb darkening tables with each new \logg, \teff, and \feh \ during the Markov chain. In addition, such a hybrid fit must be used with care, as discussed in \S\ref{sec:hybrid}.

These relations trivially simplify in the case of linear limb darkening, when $u_2=0$:

\begin{equation}
  \begin{split}
    \label{eq:linlddef}
    x_0 &= 1-\lambda^{e} \\
    x_1 &= 2/3(\lambda^{e} - \Theta(p-z)) - \lambda^{d} \\
    c_0 &= F_0 \\
    c_1 &= F_0\frac{u_1}{1 - u_1/3} \\
  \end{split}
\end{equation}

And then,

\begin{equation}
  u_1 = \frac{c_1}{c_0 + c_1/3}
\end{equation}

Given the same form ($x_0$ and $x_1$ are the same in the quadratic and linear cases), one could even fit both laws to see if the improvement in \chisq \ justifies the addition of the extra parameter.
\section{EXOFAST\_OCCULTQUAD}
\label{sec:occultquad}

The original IDL code to analytically calculate the flux decrement during transit presented in \citet{mandel02} was more than an order of magnitude faster than the previous numerical method \citep{charbonneau00, henry00}. However, the IDL version was a simple line by line translation of the Fortran code, which does not take advantage of the fact that IDL is a vector language. In our code, {\tt EXOFAST\_OCCULTQUAD}, we sped it up by a factor of 100-500 and fixed many bugs.

\subsection{Speed Enhancements}

The majority of the speed enhancement came from a major restructuring of the code to take advantage of IDL's much faster vector operations and careful attention to optimal case ordering.

With the aid of the IDL profiler, we determined the largest remaining bottleneck was in the calculation of the elliptic integral of the third kind, which used translations of the numerical recipes routines {\tt rj}, {\tt rc}, and {\tt rk}. After testing potential replacements, including IDL's separately-licensed routines {\tt IMSL\_ELRC}, {\tt IMSL\_ELRJ}, and {\tt IMSL\_RLRD}, we replaced the numerical recipes routines with an IDL translation of an ALGOL program published by \citet{bulirsch65a,bulirsch65b}. This iterative routine is many times faster and more robust than any other alternative we tried.

For our last minor speed enhancement, we combined the routines to calculate Hasting's polynomial approximation for the elliptic integrals of the first ($K(k)$) and second ($E(k)$) kind, which now share the computationally expensive task of computing $\log(1-k)$. We also tested other replacements for this routine, including Fukushima's piecewise polynomial approximation, \citep{fukushima09}, but none was reliably faster.

The calculation of the elliptic integral of the third kind is still the primary bottleneck, but is now comparable to the calculation of the elliptic integrals of the first and second kind, the arc cosine, logarithm, and IDL's {\tt WHERE} function, so additional substantive speed gains will be difficult without a fundamental re-characterization of the problem.

These speed enhancements combined make this IDL routine $\sim100$ times faster than the previous IDL version. Since this calculation is a significant fraction of calculating a transit model, this improves the run time of entire program by a significant factor. We rewrote the Fortran routine with the latter two enhancements, which is now twice as fast as its predecessor, and 2-5x faster than the current IDL version, depending on the compiler.

Lastly, we include an IDL wrapper for the Fortran version, which uses {\tt CALL\_EXTERNAL} and is a drop-in replacement for the strictly IDL version described above. After its first use, it is indistinguishable from the native Fortran in terms of speed, and is therefore 200-500 times faster than the original IDL version, and 2-5 times faster than the current IDL code, depending on Fortran the compiler used.

The downside to this wrapper is that the syntax required by IDL is technically not legal Fortran. The makefile in the IDL example does not implement f77 on Linux, likely because of compilation warnings. Some compilers, like gfortran and f95 simply refuse to compile it. It can be compiled with g77 or f77, and the warnings they issue can be safely suppressed. Only ifort will compile it without warnings. This {\tt CALL\_EXTERNAL} version has only been tested on a 32 bit Linux machine.

For those that wish to use a high-level, but non-proprietary language, we include a Python version of {\tt EXOFAST\_OCCULTQUAD}, which uses NumPy and mirrors the vector structure of the IDL version. Therefore, we expected this to be similar in execution time to the IDL, but it was actually 5 times slower. This may be a general statement about Python, or simply a result of the fact that its authors are expert IDL programmers and only novice Python programmers.

\subsection{Bug fixes and conceptual enhancements}
\label{sec:bugfix}

In addition to being relatively slow, the numerical recipes codes to calculate the elliptic integral of the third kind ({\tt rc}, {\tt rj}, {\tt rk}) were poorly-behaved where values of $z$ were within $10^{-7}$ of special cases (first, second, third, and fourth contact). Very near special cases ($10^{-13}$), the previous codes differed from the true value by as much as $10\%$. With the \citet{bulirsch65a,bulirsch65b} algorithm to calculate the elliptic integral of the third kind, the differences between the calculated value and true value (calculated with arbitrary precision in Mathematica) are less than $10^{-5}$. Since any model that suffered from this bug would be unfairly disfavored, the practical implication of this bug is the potential for a bias in the measured times that pushes data points away from these special cases. For a typical Hot Jupiter, $10^{-7}$ corresponds to $\sim90$ ms in the planetary orbit. Fortunately, this is beyond the current precision attained to date. However, diabolical alignments of several data points may further skew the inferred value.

We also fixed typos in cases where $p>1$, as in secondary eclipses. While these typos would have caused catastrophic failures, the behavior of this bug is sufficiently non-physical they would have been immediately found, so were unlikely to have done any harm. We also fixed a handful of other minor bugs, such as consistent use of double precision values and (in Fortran) functions.

The major conceptual enhancements were mentioned previously. Namely, we now allow negative values of $p$, as discussed in \S\ref{sec:otherbiases} and we return the coefficients required to analytically fit the quadratic limb darkening parameters as discussed in Appendix~\ref{sec:analyticld}.

Thorough testing at every special case, $\pm 10^{-13}$ of every special case, and for a thousand uniformly spaced values of z between $z=0$ and $z=2(1+p)$ for hundreds of values of p ranging from $10^{-13}$ to 2 have been checked, and no other errors were found.

\section{Random Number Generator}
\label{sec:random}

IDL's help states that {\tt RANDOMU}, their built-in random number generator, is similar to Numerical Recipes {\tt ran1}, which has a periodicity, m, of $\sim10^9$, and should suffice for a series of random numbers of length, $n$, such that $n \lesssim m/20 \approx 50$ million.

In our sample fit of HAT-P-3b, we generate 16 random numbers per step: one for each of the 13 parameters, two to choose the random chains, and one to compare to the likelihood ratio. We construct 26 simultaneous chains, each with a maximum of 100,000 steps. This means that in total, we may generate up to 41.6 million random numbers during the course of the fit. While this is just below the recommended number, with a few more free parameters (recall that we fixed the slope to zero and forced the orbit to be circular) or thinning, we could easily exceed the periodicity of the generator.

Initially, we were incredulous that the results could be substantively affected by the periodicity of the generator, given that the chains wander around in N-dimensional parameter space and so are unlikely to take the same step from the same place. Therefore, we constructed another random number generator which cycled through the first 10007 (a prime number) random deviates generated by {\tt RANDOMU} to artificially reduce its periodicity and more readily test this effect. While the acceptance rate was ideal, the chains were well-mixed according to our criteria, and the resulting probability distribution functions looked acceptable by eye (i.e., no obvious signs of any malfunction), the median parameters differed significantly (in some cases by more than $1.5\sigma$) from the results using {\tt RANDOMU}. Indeed, it has been shown numerous times that the quality of the random number generator can affect the scientific conclusions in other applications \citep[e.g.,][]{kalle84, parisi85, filk85, grassberger93}.

While looking for a suitable replacement, we found that the third edition of Numerical Recipes \citep{press07} warns ``be cautious about any source earlier than about 1995, since the field progressed enormously in the following decade.'' They also state that one should never use a random number generator with a periodicity of less than $\sim2 \times 10^{19}$. This certainly excludes {\tt ran1}, and even excludes their own popularly-used alternative {\tt ran2}, published in 1992, with a periodicity of $10^{18}$.

Numerical Recipes recommends a specific implementation of a random number generator, which was translated into IDL by Paolo Grigis and included with our distribution as {\tt PG\_RAN}. It has a periodicity of $\sim10^{57}$. Our program, {\tt EXOFAST\_RANDOM}, is a wrapper for this routine to allow it to return an 8-dimensional array of uniform and Gaussian random numbers in order to make it a drop-in replacement for {\tt RANDOMU}. While this version is $\sim120$ times slower than {\tt RANDOMU} because random number generators are serial by nature, and so not ideally implemented in IDL, its contribution to the overall runtime of the program is only about 3\%. A fast, built-in random number generator would essentially eliminate this contribution completely.

We repeated the fit of HAT-P-3b again, now using this generator. Because we did not approach the periodicity of {\tt RANDOMU} in this fit, it is comforting that we found no statistical difference. Further, we first did this test before implementing DE-MC\@. Repeating this test afterward, we found the effect is much less significant when using DE-MC, but still noticeable, than in a standard MCMC implementation because it is so much more efficient and therefore we take a small fraction of the maximum number of steps. In addition, the steps we do take are dominated by the two random numbers that choose the chains, not the uniform random deviate for each parameter. For generality, and because the overall speed hit is relatively small, we always use our own, slower generator. By setting a simple flag, the user can override this default behavior to select {\tt RANDOMU} or any other generator, as long as it has the same calling structure and has the ability to return 1, 2, and 3 dimensional uniform and Gaussian random deviates.

\section{Negative Eccentricity}
\label{sec:nege}

When investigating eccentricity parameterizations, we explored one that, to the best of our knowledge, has not been tried before: allowing the eccentricity to be bounded between -1 and 1. While formally undefined, we were motivated to allow a negative eccentricity by considering its definition, $e=(r_a-r_p)/(r_a+r_p)$, where $r_a$ is the distance to the focus at apoapsis and $r_p$ is the distance to the same focus at periapsis. Therefore, a negative eccentricity implies that we have incorrectly labeled our periapsis and apoapsis. When $e < 0$, we redefine a new eccentricity, $e'=-e$. Since that flips the periapsis and apoapsis, we also need to change the argument of periastron accordingly: $\omega_*' = \omega_* + \pi$. Since changing the argument of periastron also shifts the time of periastron, we need to make sure we calculate the \tp \ after changing $\omega_*$. In the end, however, we cannot allow negative eccentricities in our final distribution because it will create a discrete degeneracy between $e$ and $\omega_*$, and $-e$ and $\omega_* + \pi$. In a properly-sampled Markov Chain, the median value of $e$ would always be 0, and for statistically significant eccentricities, the chain is likely to get stuck at either the negative or positive eccentricity and never sample the other. We can avoid this by explicitly changing our values of $e$ and $\omega_*$, following the above prescription, as soon as $e$ steps to a negative value. Further, if $\omega_*$ is outside of the normally allowed bounds, $-\pi < \omega_* \leq \pi$, we can simply add or subtract $2\pi$ until it is within bounds. This scheme does not affect our uniform prior in either $e$ or $\omega_*$. By eliminating regions of parameter space that are out of bounds and therefore always rejected, the autocorrelations are smaller and the chains converge faster. Unfortunately, because we must rescale the eccentricity, it does not get rid of the Lucy-Sweeney bias as we had originally hoped.

The resultant prior distributions and a posteriori distributions were indistinguishable from the \secosw, \sesinw \ parameterization. When using a standard Markov Chain implementation and for small eccentricities, we found this parameterization to be about 20\% faster than any of those discussed in \S\ref{sec:eparam}. With the DE-MC code, it was about 20\% faster than stepping directly in $e$, where $0 \leq e < 1$, but still nearly half as fast as parametrizing it in terms of \ecosw \ and \esinw or \secosw \ and \sesinw. So while we generally recommend a DE-MC stepping in \secosw \ and \sesinw, we still present this alternative parameterization in case others find it useful.

One useful application of the same idea would be when incorporating the RM effect, where the projected rotational velocity, \vsini \ is analogous to eccentricity and the projected spin-orbit alignment, $\lambda$ is analagous to the argument of periastron. Therefore, we can allow a negative \vsini, but when it is negative, take its opposite and add $\pi$ to $\lambda$.
\section{Performance}
\label{sec:runtime}

We ran a typical case of a fit of a simulated data set with both transit and RV data with an eccentric orbit, which was well-mixed in about 5 minutes. During this time, we ran IDL's profiler to analyze the performance to see which of our routines may be a bottleneck. While the particular mix will necessarily depend on the number of data points and the details of the model, it is instructive to look at this case study in depth.

The solution to Kepler's equation is the largest remaining bottleneck, consuming $28\%$ of the total computation time. Since this iterative solution is necessarily serial, re-writing it in Fortran or C and calling it via {\tt CALL\_EXTERNAL} may be a quick way of improving the run time. However, for low-eccentricity orbits, the number of iterations is small, and for circular orbits, this calculation is trivial.

At 20\%, the next largest contribution is from the program that generates the parameter distribution and covariance plots, the vast majority going toward the later. Because we generate a plot for every parameter covariant with every other parameter, including all derived parameters, this represents over 1000 different plots. For each one, it bins every point in every chain into a 2D grid and finds the probability contours. In general, this is a useful diagnostic, but not necessary, and can be skipped by setting the appropriate flag.

The next largest contributor is the conversion to the target reference frame ($12\%$). It is also an iterative function, and is calculated twice at each step in the MCMC chain -- one for RV and one for photometry. It is important to know that this routine calls {\tt EXOFAST\_KEPLEREQ}, and its contribution is not included in that figure (so as not to double count it). The total run time of this calculation is $32\%$. This is a hefty price to pay to include an effect that is nearly negligible, and this could be skipped altogether (as has been done to date) without any significant difference. Analyzing the run time excluding this routine does not qualitatively change the top contributors.

At 6.3\%, the interpolation of the quadratic limb darkening parameters takes the next biggest slice. The tables are large, and simply loading them into memory takes a long time, though we make use of global variables to make this quicker than it normally would be.

The last major contributor, at 6.1\%, is the calculation of the transit light curve (using the all-IDL version). Our efforts to optimize this calculation are described in detail in Appendix~\ref{sec:occultquad}. Like the conversion to the target reference frame, this figure counts the time only spent inside the routine. The total contribution of this routine is 12\%. Had we left the routine alone, the total run time of the MCMC fit would have been completely dominated by this calculation, taking an hour by itself. We also note that we have not optimized the code to calculate the transit flux using a non-linear limb darkened stellar model, which was many times slower than the original code. Therefore, while in principle, it would be easy to modify our code to include non-linear limb darkening, it would increase the total run time of the fit by more than an order of magnitude.

Using our comparison between {\tt EXOFAST\_OCCULTQUAD} and its Fortran counterpart as a guide, we can guess that an all-Fortran version of EXOFAST would probably be around 5 times faster. Further, considering the execution time and relative contribution of the Fortran version of {\tt EXOFAST\_OCCULTQUAD}, an improvement of greater than 50 times is unlikely, even if all other contributions became negligible. Recently, however, NVIDIA Graphics Processing Units (GPUs) have been used with CUDA, the Compute Unified Device Architecture on highly parallelizable code for a significant (factors of 10-100) gains over their CPU counterparts \citep[e.g.][]{ford09}. While the links in a DE-MC chain are serial, each model calculation within a link and each parallel chain could be computed simultaneously to take advantage of the GPU architecture if we had enough data points to justify the GPU overheads. Indeed, preliminary tests have shown a factor of 50 improvement for the calculation of the model light curve for Kepler-sized data sets with a relatively inexpensive graphics card (GeForce GTX 480, \$200 in 2012) and show the potential for even greater improvements. However, other bottlenecks currently make the overall boost a small fraction of that, making it tough to justify additional hardware. With some effort, these other bottlenecks may also be reduced, making GPUs a promising avenue for large data sets.

Possibly the easiest way of substantively improving the run time is to decrease the total number of steps required for convergence, and therefore require fewer model calculations. Recently, \citet{mackey12} suggested an alternative way of stepping, which is worth investigating, though its advantage over DE-MC is unclear. To stress how much of a difference this may make, because of the very strong correlation between $a/R_*$ and \logg, a standard implementation of the MCMC algorithm converged nearly 1000 times slower than its DE-MC counterpart -- the difference between 3 minutes and 2 days.


\begin{thebibliography}{115}
\expandafter\ifx\csname natexlab\endcsname\relax\def\natexlab#1{#1}\fi


\bibitem[{{Agol} {et~al.}(2005){Agol}, {Steffen}, {Sari}, \&
  {Clarkson}}]{agol05}
{Agol}, E., {Steffen}, J., {Sari}, R., \& {Clarkson}, W. 2005, \mnras, 359, 567


\bibitem[{{Alonso} {et~al.}(2004){Alonso}, {Brown}, {Torres}, {Latham},
  {Sozzetti}, {Mandushev}, {Belmonte}, {Charbonneau}, {Deeg}, {Dunham},
  {O'Donovan}, \& {Stefanik}}]{alonso04}
{Alonso}, R., {et~al.} 2004, \apjl, 613, L153


\bibitem[{{Alsubai} {et~al.}(2011){Alsubai}, {Parley}, {Bramich}, {West},
  {Sorensen}, {Collier Cameron}, {Latham}, {Horne}, {Anderson}, {Bakos},
  {Brown}, {Buchhave}, {Esquerdo}, {Everett}, {F{\.z}r{\'e}sz}, {Hartman},
  {Hellier}, {Miller}, {Pollacco}, {Quinn}, {Smith}, {Stefanik}, \&
  {Szentgyorgyi}}]{alsubai11}
{Alsubai}, K.~A., {et~al.} 2011, \mnras, 417, 709


\bibitem[{{An} {et~al.}(2002){An}, {Albrow}, {Beaulieu}, {Caldwell}, {DePoy},
  {Dominik}, {Gaudi}, {Gould}, {Greenhill}, {Hill}, {Kane}, {Martin},
  {Menzies}, {Pogge}, {Pollard}, {Sackett}, {Sahu}, {Vermaak}, {Watson}, \&
  {Williams}}]{an02}
{An}, J.~H., {et~al.} 2002, \apj, 572, 521


\bibitem[{{Anderson} {et~al.}(2011){Anderson}, {Collier Cameron}, {Hellier},
  {Lendl}, {Maxted}, {Pollacco}, {Queloz}, {Smalley}, {Smith}, {Todd},
  {Triaud}, {West}, {Barros}, {Enoch}, {Gillon}, {Lister}, {Pepe},
  {S{\'e}gransan}, {Street}, \& {Udry}}]{anderson11}
{Anderson}, D.~R., {et~al.} 2011, \apjl, 726, L19


\bibitem[{{Baglin} {et~al.}(2006){Baglin}, {Auvergne}, {Boisnard}, {Lam-Trong},
  {Barge}, {Catala}, {Deleuil}, {Michel}, \& {Weiss}}]{baglin06}
{Baglin}, A., {et~al.} 2006, in COSPAR, Plenary Meeting, Vol.~36, 36th COSPAR
  Scientific Assembly, 3749--+


\bibitem[{{Bakos} {et~al.}(2004){Bakos}, {Noyes}, {Kov{\'a}cs}, {Stanek},
  {Sasselov}, \& {Domsa}}]{bakos04}
{Bakos}, G., {Noyes}, R.~W., {Kov{\'a}cs}, G., {Stanek}, K.~Z., {Sasselov},
  D.~D., \& {Domsa}, I. 2004, \pasp, 116, 266


\bibitem[{{Bakos} {et~al.}(2012){Bakos}, {Hartman}, {Torres}, {B{\'e}ky},
  {Latham}, {Buchhave}, {Csubry}, {Kov{\'a}cs}, {Bieryla}, {Quinn},
  {Szklen{\'a}r}, {Esquerdo}, {Shporer}, {Noyes}, {Fischer}, {Johnson},
  {Howard}, {Marcy}, {Sato}, {Penev}, {Everett}, {Sasselov}, {F{\H u}r{\'e}sz},
  {Stefanik}, {L{\'a}z{\'a}r}, {Papp}, \& {S{\'a}ri}}]{bakos12}
{Bakos}, G.~{\'A}., {et~al.} 2012, \aj, 144, 19


\bibitem[{{Batalha} {et~al.}(2012){Batalha}, {Rowe}, {Bryson}, {Barclay},
  {Burke}, {Caldwell}, {Christiansen}, {Mullally}, {Thompson}, {Brown},
  {Dupree}, {Fabrycky}, {Ford}, {Fortney}, {Gilliland}, {Isaacson}, {Latham},
  {Marcy}, {Quinn}, {Ragozzine}, {Shporer}, {Borucki}, {Ciardi}, {Gautier},
  {Haas}, {Jenkins}, {Koch}, {Lissauer}, {Rapin}, {Basri}, {Boss}, {Buchhave},
  {Charbonneau}, {Christensen-Dalsgaard}, {Clarke}, {Cochran}, {Demory},
  {Devore}, {Esquerdo}, {Everett}, {Fressin}, {Geary}, {Girouard}, {Gould},
  {Hall}, {Holman}, {Howard}, {Howell}, {Ibrahim}, {Kinemuchi}, {Kjeldsen},
  {Klaus}, {Li}, {Lucas}, {Morris}, {Prsa}, {Quintana}, {Sanderfer},
  {Sasselov}, {Seader}, {Smith}, {Steffen}, {Still}, {Stumpe}, {Tarter},
  {Tenenbaum}, {Torres}, {Twicken}, {Uddin}, {Van Cleve}, {Walkowicz}, \&
  {Welsh}}]{batalha12}
{Batalha}, N.~M., {et~al.} 2012, arXiv:1202.5852


\bibitem[{{Beatty} {et~al.}(2012){Beatty}, {Pepper}, {Siverd}, {Eastman},
  {Bieryla}, {Latham}, {Buchhave}, {Jensen}, {Manner}, {Stassun}, {Gaudi},
  {Berlind}, {Calkins}, {Collins}, {DePoy}, {Esquerdo}, {Fulton}, {F{\H
  u}r{\'e}sz}, {Geary}, {Gould}, {Hebb}, {Kielkopf}, {Marshall}, {Pogge},
  {Stanek}, {Stefanik}, {Street}, {Szentgyorgyi}, {Trueblood}, {Trueblood}, \&
  {Stutz}}]{beatty12}
{Beatty}, T.~G., {et~al.} 2012, \apjl, 756, L39


\bibitem[{{Borucki} {et~al.}(2010){Borucki}, {Koch}, {Basri}, {Batalha},
  {Brown}, {Caldwell}, {Caldwell}, {Christensen-Dalsgaard}, {Cochran},
  {DeVore}, {Dunham}, {Dupree}, {Gautier}, {Geary}, {Gilliland}, {Gould},
  {Howell}, {Jenkins}, {Kondo}, {Latham}, {Marcy}, {Meibom}, {Kjeldsen},
  {Lissauer}, {Monet}, {Morrison}, {Sasselov}, {Tarter}, {Boss}, {Brownlee},
  {Owen}, {Buzasi}, {Charbonneau}, {Doyle}, {Fortney}, {Ford}, {Holman},
  {Seager}, {Steffen}, {Welsh}, {Rowe}, {Anderson}, {Buchhave}, {Ciardi},
  {Walkowicz}, {Sherry}, {Horch}, {Isaacson}, {Everett}, {Fischer}, {Torres},
  {Johnson}, {Endl}, {MacQueen}, {Bryson}, {Dotson}, {Haas}, {Kolodziejczak},
  {Van Cleve}, {Chandrasekaran}, {Twicken}, {Quintana}, {Clarke}, {Allen},
  {Li}, {Wu}, {Tenenbaum}, {Verner}, {Bruhweiler}, {Barnes}, \&
  {Prsa}}]{borucki10c}
{Borucki}, W.~J., {et~al.} 2010, Science, 327, 977


\bibitem[{{Brown} {et~al.}(2001){Brown}, {Charbonneau}, {Gilliland}, {Noyes},
  \& {Burrows}}]{brown01}
{Brown}, T.~M., {Charbonneau}, D., {Gilliland}, R.~L., {Noyes}, R.~W., \&
  {Burrows}, A. 2001, \apj, 552, 699


\bibitem[{{Bulirsch}(1965{\natexlab{a}})}]{bulirsch65a}
{Burlirsch}, R. 1965{\natexlab{a}}, Numerische Mathematik, 7, 78


\bibitem[{{Bulirsch}(1965{\natexlab{b}})}]{bulirsch65b}
---. 1965{\natexlab{b}}, Numerische Mathematik, 7, 353


\bibitem[{{Carter} \& {Winn}(2009)}]{carter09}
{Carter}, J.~A., \& {Winn}, J.~N. 2009, \apj, 704, 51


\bibitem[{{Carter} \& {Winn}(2010)}]{carter10}
---. 2010, \apj, 716, 850


\bibitem[{{Carter} {et~al.}(2011{\natexlab{a}}){Carter}, {Winn}, {Holman},
  {Fabrycky}, {Berta}, {Burke}, \& {Nutzman}}]{carter11}
{Carter}, J.~A., {Winn}, J.~N., {Holman}, M.~J., {Fabrycky}, D., {Berta},
  Z.~K., {Burke}, C.~J., \& {Nutzman}, P. 2011{\natexlab{a}}, \apj, 730, 82


\bibitem[{{Carter} {et~al.}(2008){Carter}, {Yee}, {Eastman}, {Gaudi}, \&
  {Winn}}]{carter08}
{Carter}, J.~A., {Yee}, J.~C., {Eastman}, J., {Gaudi}, B.~S., \& {Winn}, J.~N.
  2008, \apj, 689, 499


\bibitem[{{Carter} {et~al.}(2011{\natexlab{b}}){Carter}, {Fabrycky},
  {Ragozzine}, {Holman}, {Quinn}, {Latham}, {Buchhave}, {Van Cleve}, {Cochran},
  {Cote}, {Endl}, {Ford}, {Haas}, {Jenkins}, {Koch}, {Li}, {Lissauer},
  {MacQueen}, {Middour}, {Orosz}, {Rowe}, {Steffen}, \& {Welsh}}]{carter11b}
{Carter}, J.~A., {et~al.} 2011{\natexlab{b}}, Science, 331, 562


\bibitem[{{Charbonneau} {et~al.}(2000){Charbonneau}, {Brown}, {Latham}, \&
  {Mayor}}]{charbonneau00}
{Charbonneau}, D., {Brown}, T.~M., {Latham}, D.~W., \& {Mayor}, M. 2000, \apjl,
  529, L45


\bibitem[{{Charbonneau} {et~al.}(2002){Charbonneau}, {Brown}, {Noyes}, \&
  {Gilliland}}]{charbonneau02}
{Charbonneau}, D., {Brown}, T.~M., {Noyes}, R.~W., \& {Gilliland}, R.~L. 2002,
  \apj, 568, 377


\bibitem[{{Charbonneau} {et~al.}(2006){Charbonneau}, {Winn}, {Latham}, {Bakos},
  {Falco}, {Holman}, {Noyes}, {Cs{\'a}k}, {Esquerdo}, {Everett}, \&
  {O'Donovan}}]{charbonneau06}
{Charbonneau}, D., {et~al.} 2006, \apj, 636, 445


\bibitem[{{Claret} \& {Bloemen}(2011)}]{claret11}
{Claret}, A., \& {Bloemen}, S. 2011, \aap, 529, A75+


\bibitem[{{Collier Cameron} {et~al.}(2007){Collier Cameron}, {Bouchy},
  {H{\'e}brard}, {Maxted}, {Pollacco}, {Pont}, {Skillen}, {Smalley}, {Street},
  {West}, {Wilson}, {Aigrain}, {Christian}, {Clarkson}, {Enoch}, {Evans},
  {Fitzsimmons}, {Fleenor}, {Gillon}, {Haswell}, {Hebb}, {Hellier}, {Hodgkin},
  {Horne}, {Irwin}, {Kane}, {Keenan}, {Loeillet}, {Lister}, {Mayor}, {Moutou},
  {Norton}, {Osborne}, {Parley}, {Queloz}, {Ryans}, {Triaud}, {Udry}, \&
  {Wheatley}}]{cameron07}
{Collier Cameron}, A., {et~al.} 2007, \mnras, 375, 951


\bibitem[{{Dawson} \& {Fabrycky}(2010)}]{dawson10}
{Dawson}, R.~I., \& {Fabrycky}, D.~C. 2010, \apj, 722, 937


\bibitem[{{Deming} {et~al.}(2006){Deming}, {Harrington}, {Seager}, \&
  {Richardson}}]{deming06}
{Deming}, D., {Harrington}, J., {Seager}, S., \& {Richardson}, L.~J. 2006,
  \apj, 644, 560


\bibitem[{{Deming} {et~al.}(2005){Deming}, {Seager}, {Richardson}, \&
  {Harrington}}]{deming05}
{Deming}, D., {Seager}, S., {Richardson}, L.~J., \& {Harrington}, J. 2005,
  \nat, 434, 740


\bibitem[{{Doyle} {et~al.}(2011){Doyle}, {Carter}, {Fabrycky}, {Slawson},
  {Howell}, {Winn}, {Orosz}, {Pr{\v}sa}, {Welsh}, {Quinn}, {Latham}, {Torres},
  {Buchhave}, {Marcy}, {Fortney}, {Shporer}, {Ford}, {Lissauer}, {Ragozzine},
  {Rucker}, {Batalha}, {Jenkins}, {Borucki}, {Koch}, {Middour}, {Hall},
  {McCauliff}, {Fanelli}, {Quintana}, {Holman}, {Caldwell}, {Still},
  {Stefanik}, {Brown}, {Esquerdo}, {Tang}, {Furesz}, {Geary}, {Berlind},
  {Calkins}, {Short}, {Steffen}, {Sasselov}, {Dunham}, {Cochran}, {Boss},
  {Haas}, {Buzasi}, \& {Fischer}}]{doyle11}
{Doyle}, L.~R., {et~al.} 2011, Science, 333, 1602


\bibitem[{{Eastman} {et~al.}(2010{\natexlab{a}}){Eastman}, {Gaudi}, {Siverd},
  {Trueblood}, \& {Trueblood}}]{eastman10a}
{Eastman}, J., {Gaudi}, B.~S., {Siverd}, R., {Trueblood}, M., \& {Trueblood},
  P. 2010{\natexlab{a}}, in Society of Photo-Optical Instrumentation Engineers
  (SPIE) Conference Series, Vol. 7733, Society of Photo-Optical Instrumentation
  Engineers (SPIE) Conference Series


\bibitem[{{Eastman} {et~al.}(2010{\natexlab{b}}){Eastman}, {Siverd}, \&
  {Gaudi}}]{eastman10b}
{Eastman}, J., {Siverd}, R., \& {Gaudi}, B.~S. 2010{\natexlab{b}}, \pasp, 122,
  935


\bibitem[{{Enoch} {et~al.}(2010){Enoch}, {Collier Cameron}, {Parley}, \&
  {Hebb}}]{enoch10}
{Enoch}, B., {Collier Cameron}, A., {Parley}, N.~R., \& {Hebb}, L. 2010, \aap,
  516, A33+


\bibitem[{{Fabrycky}(2008)}]{fabrycky08}
{Fabrycky}, D.~C. 2008, in AAS/Division of Dynamical Astronomy Meeting,
  Vol.~39, AAS/Division of Dynamical Astronomy Meeting \#39, \#06.07--+


\bibitem[{Filk {et~al.}(1985)Filk, Marcu, \& Fredenhagen}]{filk85}
Filk, T., Marcu, M., \& Fredenhagen, K. 1985, Physics Letters B, 165, 125


\bibitem[{{Fleming} {et~al.}(2010){Fleming}, {Ge}, {Mahadevan}, {Lee},
  {Eastman}, {Siverd}, {Gaudi}, {Niedzielski}, {Sivarani}, {Stassun},
  {Wolszczan}, {Barnes}, {Gary}, {Cuong Nguyen}, {Morehead}, {Wan}, {Zhao},
  {Liu}, {Guo}, {Kane}, {van Eyken}, {De Lee}, {Crepp}, {Shelden}, {Laws},
  {Wisniewski}, {Schneider}, {Pepper}, {Snedden}, {Pan}, {Bizyaev},
  {Brewington}, {Malanushenko}, {Malanushenko}, {Oravetz}, {Simmons}, \&
  {Watters}}]{fleming10}
{Fleming}, S.~W., {et~al.} 2010, \apj, 718, 1186


\bibitem[{{Ford}(2005)}]{ford05}
{Ford}, E.~B. 2005, \aj, 129, 1706


\bibitem[{{Ford}(2006)}]{ford06a}
---. 2006, \apj, 642, 505


\bibitem[{{Ford}(2009)}]{ford09}
---. 2009, \na, 14, 406


\bibitem[{{Ford} \& {Gaudi}(2006)}]{ford06b}
{Ford}, E.~B., \& {Gaudi}, B.~S. 2006, \apjl, 652, L137


\bibitem[{{Ford} \& {Holman}(2007)}]{ford07}
{Ford}, E.~B., \& {Holman}, M.~J. 2007, \apjl, 664, L51


\bibitem[{{Foreman-Mackey} {et~al.}(2012){Foreman-Mackey}, {Hogg}, {Lang}, \&
  {Goodman}}]{mackey12}
{Foreman-Mackey}, D., {Hogg}, D.~W., {Lang}, D., \& {Goodman}, J. 2012, arXiv:1202.3665


\bibitem[{{Fortney} {et~al.}(2006){Fortney}, {Saumon}, {Marley}, {Lodders}, \&
  {Freedman}}]{fortney06}
{Fortney}, J.~J., {Saumon}, D., {Marley}, M.~S., {Lodders}, K., \& {Freedman},
  R.~S. 2006, \apj, 642, 495


\bibitem[{{Fukushima}(2009)}]{fukushima09}
{Fukushima}, T. 2009, Celestial Mechanics and Dynamical Astronomy, 105, 305


\bibitem[{{Gaudi} \& {Winn}(2007)}]{gaudi07}
{Gaudi}, B.~S., \& {Winn}, J.~N. 2007, \apj, 655, 550


\bibitem[{{Gazak} {et~al.}(2012){Gazak}, {Johnson}, {Tonry}, {Dragomir},
  {Eastman}, {Mann}, \& {Agol}}]{gazak11}
{Gazak}, J.~Z., {Johnson}, J.~A., {Tonry}, J., {Dragomir}, D., {Eastman}, J.,
  {Mann}, A.~W., \& {Agol}, E. 2012, Advances in Astronomy, 2012


\bibitem[{{Gelman} {et~al.}(2003){Gelman}, {Carlin}, {Stern}, \&
  {Rubin}}]{gelman03}
{Gelman}, A., {Carlin}, J.~B., {Stern}, H.~S., \& {Rubin}, D.~B. 2003,
  {Bayesian Data Analysis}, 2nd edn.


\bibitem[{{Gould}(2003)}]{gould03}
{Gould}, A. 2003, arXiv:0310577


\bibitem[{Grassberger(1993)}]{grassberger93}
Grassberger, P. 1993, Physics Letters A, 181, 43


\bibitem[{{Gregory}(2005)}]{gregory05}
{Gregory}, P.~C. 2005, \apj, 631, 1198


\bibitem[{{Guillot}(2005)}]{guillot05}
{Guillot}, T. 2005, Annual Review of Earth and Planetary Sciences, 33, 493


\bibitem[{{Hartman} {et~al.}(2008){Hartman}, {Gaudi}, {Holman}, {McLeod},
  {Stanek}, {Barranco}, {Pinsonneault}, \& {Kalirai}}]{hartman08}
{Hartman}, J.~D., {Gaudi}, B.~S., {Holman}, M.~J., {McLeod}, B.~A., {Stanek},
  K.~Z., {Barranco}, J.~A., {Pinsonneault}, M.~H., \& {Kalirai}, J.~S. 2008,
  \apj, 675, 1254


\bibitem[{{Hayek} {et~al.}(2012){Hayek}, {Sing}, {Pont}, \&
  {Asplund}}]{hayek12}
{Hayek}, W., {Sing}, D., {Pont}, F., \& {Asplund}, M. 2012, \aap, 539, A102


\bibitem[{{Hellier} {et~al.}(2009){Hellier}, {Anderson}, {Cameron}, {Gillon},
  {Hebb}, {Maxted}, {Queloz}, {Smalley}, {Triaud}, {West}, {Wilson}, {Bentley},
  {Enoch}, {Horne}, {Irwin}, {Lister}, {Mayor}, {Parley}, {Pepe}, {Pollacco},
  {Segransan}, {Udry}, \& {Wheatley}}]{hellier09}
{Hellier}, C., {et~al.} 2009, \nat, 460, 1098


\bibitem[{{Henry} {et~al.}(2000){Henry}, {Marcy}, {Butler}, \&
  {Vogt}}]{henry00}
{Henry}, G.~W., {Marcy}, G.~W., {Butler}, R.~P., \& {Vogt}, S.~S. 2000, \apjl,
  529, L41


\bibitem[{{Holman} \& {Murray}(2005)}]{holman05}
{Holman}, M.~J., \& {Murray}, N.~W. 2005, Science, 307, 1288


\bibitem[{{Johnson} {et~al.}(2011){Johnson}, {Payne}, {Howard}, {Clubb},
  {Ford}, {Bowler}, {Henry}, {Fischer}, {Marcy}, {Brewer}, {Schwab}, {Reffert},
  \& {Lowe}}]{johnson11}
{Johnson}, J.~A., {et~al.} 2011, \aj, 141, 16


\bibitem[{{Jord{\'a}n} \& {Bakos}(2008)}]{jordan08}
{Jord{\'a}n}, A., \& {Bakos}, G.~{\'A}. 2008, \apj, 685, 543


\bibitem[{Kalle \& Wansleben(1984)}]{kalle84}
Kalle, C., \& Wansleben, S. 1984, Computer Physics Communications, 33, 343


\bibitem[{{Klinglesmith} \& {Sobieski}(1970)}]{klinglesmith70}
{Klinglesmith}, D.~A., \& {Sobieski}, S. 1970, \aj, 75, 175


\bibitem[{{Knutson} {et~al.}(2009){Knutson}, {Charbonneau}, {Cowan}, {Fortney},
  {Showman}, {Agol}, \& {Henry}}]{knutson09}
{Knutson}, H.~A., {Charbonneau}, D., {Cowan}, N.~B., {Fortney}, J.~J.,
  {Showman}, A.~P., {Agol}, E., \& {Henry}, G.~W. 2009, \apj, 703, 769


\bibitem[{{Knutson} {et~al.}(2007){Knutson}, {Charbonneau}, {Allen}, {Fortney},
  {Agol}, {Cowan}, {Showman}, {Cooper}, \& {Megeath}}]{knutson07}
{Knutson}, H.~A., {et~al.} 2007, \nat, 447, 183


\bibitem[{{Koch} {et~al.}(2010){Koch}, {Borucki}, {Basri}, {Batalha}, {Brown},
  {Caldwell}, {Christensen-Dalsgaard}, {Cochran}, {DeVore}, {Dunham},
  {Gautier}, {Geary}, {Gilliland}, {Gould}, {Jenkins}, {Kondo}, {Latham},
  {Lissauer}, {Marcy}, {Monet}, {Sasselov}, {Boss}, {Brownlee}, {Caldwell},
  {Dupree}, {Howell}, {Kjeldsen}, {Meibom}, {Morrison}, {Owen}, {Reitsema},
  {Tarter}, {Bryson}, {Dotson}, {Gazis}, {Haas}, {Kolodziejczak}, {Rowe}, {Van
  Cleve}, {Allen}, {Chandrasekaran}, {Clarke}, {Li}, {Quintana}, {Tenenbaum},
  {Twicken}, \& {Wu}}]{koch10b}
{Koch}, D.~G., {et~al.} 2010, \apjl, 713, L79


\bibitem[{{Kov{\'a}cs} {et~al.}(2002){Kov{\'a}cs}, {Zucker}, \&
  {Mazeh}}]{kovacs02}
{Kov{\'a}cs}, G., {Zucker}, S., \& {Mazeh}, T. 2002, \aap, 391, 369


\bibitem[{{Landsman}(1993)}]{landsman93}
{Landsman}, W.~B. 1993, in Astronomical Society of the Pacific Conference
  Series, Vol.~52, Astronomical Data Analysis Software and Systems II, ed.
  {R.~J.~Hanisch, R.~J.~V.~Brissenden, \& J.~Barnes}, 246--+


\bibitem[{{Laughlin} {et~al.}(2005){Laughlin}, {Marcy}, {Vogt}, {Fischer}, \&
  {Butler}}]{laughlin05}
{Laughlin}, G., {Marcy}, G.~W., {Vogt}, S.~S., {Fischer}, D.~A., \& {Butler},
  R.~P. 2005, \apjl, 629, L121


\bibitem[{{Lee} {et~al.}(2011){Lee}, {Ge}, {Fleming}, {Stassun}, {Gaudi},
  {Barnes}, {Mahadevan}, {Eastman}, {Wright}, {Siverd}, {Gary}, {Ghezzi},
  {Laws}, {Wisniewski}, {Porto de Mello}, {Ogando}, {Maia}, {Nicolaci da
  Costa}, {Sivarani}, {Pepper}, {Cuong Nguyen}, {Hebb}, {De Lee}, {Wang},
  {Wan}, {Zhao}, {Chang}, {Groot}, {Varosi}, {Hearty}, {Hanna}, {van Eyken},
  {Kane}, {Agol}, {Bizyaev}, {Bochanski}, {Brewington}, {Chen}, {Costello},
  {Dou}, {Eisenstein}, {Fletcher}, {Ford}, {Guo}, {Holtzman}, {Jiang}, {French
  Leger}, {Liu}, {Long}, {Malanushenko}, {Malanushenko}, {Malik}, {Oravetz},
  {Pan}, {Rohan}, {Schneider}, {Shelden}, {Snedden}, {Simmons}, {Weaver},
  {Weinberg}, \& {Xie}}]{lee11}
{Lee}, B.~L., {et~al.} 2011, \apj, 728, 32


\bibitem[{Levenberg(1944)}]{levenberg44}
Levenberg, K. 1944, Quarterly Journal of Applied Mathmatics, II, 164


\bibitem[{{Loeb}(2005)}]{loeb05}
{Loeb}, A. 2005, \apjl, 623, L45


\bibitem[{{Lomb}(1976)}]{lomb76}
{Lomb}, N.~R. 1976, \apss, 39, 447


\bibitem[{{Lucy}(2012)}]{lucy12}
{Lucy}, L.~B. 2012, arXiv:1206.3462


\bibitem[{{Lucy} \& {Sweeney}(1971)}]{lucy71}
{Lucy}, L.~B., \& {Sweeney}, M.~A. 1971, \aj, 76, 544


\bibitem[{{Lutz} \& {Kelker}(1973)}]{lutz73}
{Lutz}, T.~E., \& {Kelker}, D.~H. 1973, \pasp, 85, 573


\bibitem[{{Mandel} \& {Agol}(2002)}]{mandel02}
{Mandel}, K., \& {Agol}, E. 2002, \apjl, 580, L171


\bibitem[{{Markwardt}(2009)}]{markwardt09}
{Markwardt}, C.~B. 2009, in Astronomical Society of the Pacific Conference
  Series, Vol. 411, Astronomical Data Analysis Software and Systems XVIII, ed.
  {D.~A.~Bohlender, D.~Durand, \& P.~Dowler}, 251


\bibitem[{Marquardt(1963)}]{marquardt63}
Marquardt, D.~W. 1963, Siam Journal on Applied Mathematics, 11


\bibitem[{{Mayor} \& {Queloz}(1995)}]{mayor95}
{Mayor}, M., \& {Queloz}, D. 1995, \nat, 378, 355


\bibitem[{{McCullough} {et~al.}(2006){McCullough}, {Stys}, {Valenti},
  {Johns-Krull}, {Janes}, {Heasley}, {Bye}, {Dodd}, {Fleming}, {Pinnick},
  {Bissinger}, {Gary}, {Howell}, \& {Vanmunster}}]{mccullough06}
{McCullough}, P.~R., {et~al.} 2006, \apj, 648, 1228


\bibitem[{{Meschiari} {et~al.}(2009){Meschiari}, {Wolf}, {Rivera}, {Laughlin},
  {Vogt}, \& {Butler}}]{meschiari09}
{Meschiari}, S., {Wolf}, A.~S., {Rivera}, E., {Laughlin}, G., {Vogt}, S., \&
  {Butler}, P. 2009, \pasp, 121, 1016


\bibitem[{{Mikkola}(1987)}]{mikkola87}
{Mikkola}, S. 1987, Celestial Mechanics, 40, 329


\bibitem[{{Miralda-Escud{\'e}}(2002)}]{miralda02}
{Miralda-Escud{\'e}}, J. 2002, \apj, 564, 1019


\bibitem[{{Mislis} {et~al.}(2012){Mislis}, {Heller}, {Fernandez}, {Seemann},
  {Ioannidis}, \& {Avdellidou}}]{mislis12}
{Mislis}, D., {Heller}, R., {Fernandez}, J., {Seemann}, U., {Ioannidis}, P., \&
  {Avdellidou}, C. 2012, in 10th Hellenic Astronomical Conference, ed.
  I.~{Papadakis} \& A.~{Anastasiadis}, 12--12


\bibitem[{Nelder \& Mead(1965)}]{nelder65}
Nelder, J.~A., \& Mead, R. 1965, The Computer Journal, 7, 308


\bibitem[{{P{\'a}l}(2012)}]{pal12}
{P{\'a}l}, A. 2012, \mnras, 421, 1825


\bibitem[{{P{\'a}l} \& {Kocsis}(2008)}]{pal08a}
{P{\'a}l}, A., \& {Kocsis}, B. 2008, \mnras, 389, 191


\bibitem[{Parisi \& Rapuano(1985)}]{parisi85}
Parisi, G., \& Rapuano, F. 1985, Physics Letters B, 157, 301


\bibitem[{{Pepper} {et~al.}(2007){Pepper}, {Pogge}, {DePoy}, {Marshall},
  {Stanek}, {Stutz}, {Poindexter}, {Siverd}, {O'Brien}, {Trueblood}, \&
  {Trueblood}}]{pepper07}
{Pepper}, J., {et~al.} 2007, \pasp, 119, 923


\bibitem[{{Pepper} {et~al.}(2012){Pepper}, {Siverd}, {Beatty}, {Gaudi},
 {Stassun}, {Eastman}, {Collins}, {Latham}, {Bieryla}, {Buchhave}, {Jensen},
 {Manner}, {Penev}, {Crepp}, {Cargile}, {Dhital}, {Calkins}, {Esquerdo},
 {Berlind}, {Fulton}, {Street}, {Mao}, {Richert}, {Gould}, {DePoy},
 {Kielkopf}, {Marshall}, {Pogge}, {Stefanik}, {Trueblood}, \&
 {Trueblood}}]{pepper12}
---. 2012, arXiv:1211.1031


\bibitem[{{Press} {et~al.}(2007){Press}, {Teukolsky}, {Vetterling}, \&
  {Flannery}}]{press07}
{Press}, W.~H., {Teukolsky}, S.~A., {Vetterling}, W.~T., \& {Flannery}, B.~P.
  2007, {Numerical recipes : the art of scientific computing}, 3rd edn., ed.
  {Press, W.~H.}


\bibitem[{{Pr{\v s}a} \& {Zwitter}(2005)}]{prsa05}
{Pr{\v s}a}, A., \& {Zwitter}, T. 2005, \apj, 628, 426


\bibitem[{{Queloz} {et~al.}(2000){Queloz}, {Eggenberger}, {Mayor}, {Perrier},
  {Beuzit}, {Naef}, {Sivan}, \& {Udry}}]{queloz00}
{Queloz}, D., {Eggenberger}, A., {Mayor}, M., {Perrier}, C., {Beuzit}, J.~L.,
  {Naef}, D., {Sivan}, J.~P., \& {Udry}, S. 2000, \aap, 359, L13


\bibitem[{{Rafikov}(2009)}]{rafikov09}
{Rafikov}, R.~R. 2009, \apj, 700, 965


\bibitem[{{Ragozzine} \& {Wolf}(2009)}]{ragozzine09}
{Ragozzine}, D., \& {Wolf}, A.~S. 2009, \apj, 698, 1778


\bibitem[{{Rauscher} {et~al.}(2007){Rauscher}, {Menou}, {Cho}, {Seager}, \&
  {Hansen}}]{rauscher07}
{Rauscher}, E., {Menou}, K., {Cho}, J.~Y.-K., {Seager}, S., \& {Hansen},
  B.~M.~S. 2007, \apjl, 662, L115


\bibitem[{{Sasselov}(2003)}]{sasselov03}
{Sasselov}, D.~D. 2003, \apj, 596, 1327


\bibitem[{{Sato} {et~al.}(2005){Sato}, {Fischer}, {Henry}, {Laughlin},
  {Butler}, {Marcy}, {Vogt}, {Bodenheimer}, {Ida}, {Toyota}, {Wolf}, {Valenti},
  {Boyd}, {Johnson}, {Wright}, {Ammons}, {Robinson}, {Strader}, {McCarthy},
  {Tah}, \& {Minniti}}]{sato05}
{Sato}, B., {et~al.} 2005, \apj, 633, 465


\bibitem[{{Scargle}(1982)}]{scargle82}
{Scargle}, J.~D. 1982, \apj, 263, 835


\bibitem[{{Scharf}(2007)}]{scharf07}
{Scharf}, C.~A. 2007, \apj, 661, 1218


\bibitem[{{Seager} \& {Mall{\'e}n-Ornelas}(2003)}]{seager03}
{Seager}, S., \& {Mall{\'e}n-Ornelas}, G. 2003, \apj, 585, 1038


\bibitem[{{Siverd} {et~al.}(2012){Siverd}, {Beatty}, {Pepper}, {Eastman},
  {Collins}, {Bieryla}, {Latham}, {Buchhave}, {Jensen}, {Crepp}, {Street},
  {Stassun}, {Gaudi}, {Berlind}, {Calkins}, {DePoy}, {Esquerdo}, {Fulton},
  {Furesz}, {Geary}, {Gould}, {Hebb}, {Kielkopf}, {Marshall}, {Pogge},
  {Stanek}, {Stefanik}, {Szentgyorgyi}, {Trueblood}, {Trueblood}, {Stutz}, \&
  {van Saders}}]{siverd12}
{Siverd}, R.~J., {et~al.} 2012, arXiv:1206.1635


\bibitem[{{Southworth}(2008)}]{southworth08}
{Southworth}, J. 2008, \mnras, 386, 1644


\bibitem[{{Sozzetti} {et~al.}(2007){Sozzetti}, {Torres}, {Charbonneau},
  {Latham}, {Holman}, {Winn}, {Laird}, \& {O'Donovan}}]{sozzetti07}
{Sozzetti}, A., {Torres}, G., {Charbonneau}, D., {Latham}, D.~W., {Holman},
  M.~J., {Winn}, J.~N., {Laird}, J.~B., \& {O'Donovan}, F.~T. 2007, \apj, 664,
  1190


\bibitem[{{Steffen} \& {Agol}(2005)}]{steffen05}
{Steffen}, J.~H., \& {Agol}, E. 2005, \mnras, 364, L96


\bibitem[{{Tegmark} {et~al.}(2004){Tegmark}, {Strauss}, {Blanton}, {Abazajian},
  {Dodelson}, {Sandvik}, {Wang}, {Weinberg}, {Zehavi}, {Bahcall}, {Hoyle},
  {Schlegel}, {Scoccimarro}, {Vogeley}, {Berlind}, {Budavari}, {Connolly},
  {Eisenstein}, {Finkbeiner}, {Frieman}, {Gunn}, {Hui}, {Jain}, {Johnston},
  {Kent}, {Lin}, {Nakajima}, {Nichol}, {Ostriker}, {Pope}, {Scranton},
  {Seljak}, {Sheth}, {Stebbins}, {Szalay}, {Szapudi}, {Xu}, {Annis},
  {Brinkmann}, {Burles}, {Castander}, {Csabai}, {Loveday}, {Doi}, {Fukugita},
  {Gillespie}, {Hennessy}, {Hogg}, {Ivezi{\'c}}, {Knapp}, {Lamb}, {Lee},
  {Lupton}, {McKay}, {Kunszt}, {Munn}, {O'Connell}, {Peoples}, {Pier},
  {Richmond}, {Rockosi}, {Schneider}, {Stoughton}, {Tucker}, {vanden Berk},
  {Yanny}, \& {York}}]{tegmark04}
{Tegmark}, M., {et~al.} 2004, \prd, 69, 103501


\bibitem[{{ter Braak}(2006)}]{braak06}
{ter Braak}, C. J.~F. 2006, Statistics and Computing, 16, 239


\bibitem[{{Torres} {et~al.}(2010){Torres}, {Andersen}, \&
  {Gim{\'e}nez}}]{torres10}
{Torres}, G., {Andersen}, J., \& {Gim{\'e}nez}, A. 2010, \aapr, 18, 67


\bibitem[{{Torres} {et~al.}(2007){Torres}, {Bakos}, {Kov{\'a}cs}, {Latham},
  {Fern{\'a}ndez}, {Noyes}, {Esquerdo}, {Sozzetti}, {Fischer}, {Butler},
  {Marcy}, {Stefanik}, {Sasselov}, {L{\'a}z{\'a}r}, {Papp}, \&
  {S{\'a}ri}}]{torres07}
{Torres}, G., {et~al.} 2007, \apjl, 666, L121


\bibitem[{{Triaud} {et~al.}(2010){Triaud}, {Collier Cameron}, {Queloz},
  {Anderson}, {Gillon}, {Hebb}, {Hellier}, {Loeillet}, {Maxted}, {Mayor},
  {Pepe}, {Pollacco}, {S{\'e}gransan}, {Smalley}, {Udry}, {West}, \&
  {Wheatley}}]{triaud10}
{Triaud}, A.~H.~M.~J., {et~al.} 2010, \aap, 524, A25


\bibitem[{{Vidal-Madjar} {et~al.}(2003){Vidal-Madjar}, {Lecavelier des Etangs},
  {D{\'e}sert}, {Ballester}, {Ferlet}, {H{\'e}brard}, \& {Mayor}}]{vidal03}
{Vidal-Madjar}, A., {Lecavelier des Etangs}, A., {D{\'e}sert}, J., {Ballester},
  G.~E., {Ferlet}, R., {H{\'e}brard}, G., \& {Mayor}, M. 2003, \nat, 422, 143


\bibitem[{{Winn}(2010)}]{winn10a}
{Winn}, J.~N. 2010, {Exoplanet Transits and Occultations}, ed. {Seager, S.},
  55--77


\bibitem[{{Winn} {et~al.}(2005){Winn}, {Noyes}, {Holman}, {Charbonneau},
  {Ohta}, {Taruya}, {Suto}, {Narita}, {Turner}, {Johnson}, {Marcy}, {Butler},
  \& {Vogt}}]{winn05}
{Winn}, J.~N., {et~al.} 2005, \apj, 631, 1215


\bibitem[{{Winn} {et~al.}(2009){Winn}, {Howard}, {Johnson}, {Marcy}, {Gazak},
  {Starkey}, {Ford}, {Col{\'o}n}, {Reyes}, {Nortmann}, {Dreizler}, {Odewahn},
  {Welsh}, {Kadakia}, {Vanderbei}, {Adams}, {Lockhart}, {Crossfield},
  {Valenti}, {Dantowitz}, \& {Carter}}]{winn09b}
---. 2009, \apj, 703, 2091


\bibitem[{{Winn} {et~al.}(2010){Winn}, {Johnson}, {Howard}, {Marcy}, {Bakos},
  {Hartman}, {Torres}, {Albrecht}, \& {Narita}}]{winn10b}
---. 2010, \apj, 718, 575


\bibitem[{{Wisniewski} {et~al.}(2012){Wisniewski}, {Ge}, {Crepp}, {De Lee},
  {Eastman}, {Esposito}, {Fleming}, {Gaudi}, {Ghezzi}, {Gonzalez Hernandez},
  {Lee}, {Stassun}, {Agol}, {Allende Prieto}, {Barnes}, {Bizyaev}, {Cargile},
  {Chang}, {Da Costa}, {Porto De Mello}, {Femen{\'{\i}}a}, {Ferreira}, {Gary},
  {Hebb}, {Holtzman}, {Liu}, {Ma}, {Mack}, {Mahadevan}, {Maia}, {Nguyen},
  {Ogando}, {Oravetz}, {Paegert}, {Pan}, {Pepper}, {Rebolo}, {Santiago},
  {Schneider}, {Shelden}, {Simmons}, {Tofflemire}, {Wan}, {Wang}, \&
  {Zhao}}]{wisniewski12}
{Wisniewski}, J.~P., {et~al.} 2012, \aj, 143, 107


\bibitem[{{Wright} \& {Howard}(2009)}]{wright09}
{Wright}, J.~T., \& {Howard}, A.~W. 2009, \apjs, 182, 205


\bibitem[{{Wright} {et~al.}(2011){Wright}, {Fakhouri}, {Marcy}, {Han}, {Feng},
  {Johnson}, {Howard}, {Fischer}, {Valenti}, {Anderson}, \&
  {Piskunov}}]{wright11}
{Wright}, J.~T., {et~al.} 2011, \pasp, 123, 412


\bibitem[{{Yi} {et~al.}(2001){Yi}, {Demarque}, {Kim}, {Lee}, {Ree}, {Lejeune},
  \& {Barnes}}]{yi01}
{Yi}, S., {Demarque}, P., {Kim}, Y.-C., {Lee}, Y.-W., {Ree}, C.~H., {Lejeune},
  T., \& {Barnes}, S. 2001, \apjs, 136, 417


\end{thebibliography}
\bibliographystyle{apj}



\end{document}